\documentclass[12pt]{article}

\usepackage{graphicx}

\begin{document}
        
\title{``PAIR PRODUCTION OF SCALAR TOP
 QUARKS IN $e^{+}e^{-}$ COLLISIONS AT ILC''}

\author{A.~Bartl$^{a,b}$, ~~ W.~Majerotto$^{c}$, \\
K.~M$\ddot o$nig$^{d}$, ~~ A.N.~Skachkova$^{e}$, 
~~  N.B.~Skachkov$^{e}$ }

\maketitle
\begin{center}
{\normalsize \it $^a$ University of Vienna,
 Faculty of Physics, 1090 Vienna,
 Boltzmanngasse 5, Austria. \\
 $^b$ AHEP Group, Instituto de Fisica
 Corpuscular - C.S.I.C.,  Universidad de Valencia, 
 Edificio Institutos de Investigacion,
 Apt. 22085, E-46071 Valencia, Spain \\
 $^c$ Institute for High Energy Physics 
 (HEPHY Vienna), Nikolsdorfergasse 18, 
 A-1050 Vienna, Austria.\\
 $^d$ DESY, Platanenallee 6,  
 D-15738 Zeuthen, Germany. \\
 $^e$ JINR, Joliot-Curie 6, 141980 Dubna,
 Moscow region, Russia. \\}
\end{center}

\bigskip
\begin{abstract}
\noindent 

    We study the  pair production of scalar top 
    quarks  $\tilde t_{1}$ in  $e^{+}e^{-}$
    collisions  with the subsequent decay 
 into  $b$-quarks and  charginos,  
    $\tilde t_{1} \to b \tilde \chi_{1}^{\pm}$. 
    We simulate this process  using PYTHIA6.4
    for  beam  energies   $2E_{beam}=\sqrt {s} = 350, 400, 500, 800, 
    1000$ GeV.  Proposing a set of criteria we 
obtain a good separation of the signal
stop events from top quark pair production
 which is the main background.
The number of stop production events obtained with the proposed cuts for
different energies is calculated for an integrated luminosity of
1000 $fb^{-1}$. We propose a method to reconstruct the mass of the top
squark, provided the mass of the lightest neutralino is known, and
estimate the error of the mass determination for the case $\sqrt{s}$ = 500
GeV.

\end{abstract}


\section{Introduction.}

 ~~~~The scalar top quark, the bosonic
 partner of the top quark, has attracted 
 much attention  as it is expected to be 
 the lightest colored 
 supersymmetric (SUSY) \cite{SUSY} particle.
 $\tilde t_{L}$ and $\tilde t_{R}$, the
 supersymmetric partners of the left-handed 
 and right-handed top quark,  mix and the
 resulting two mass eigenstates 
 $\tilde t_{1}$ and $\tilde t_{2}$, can	 
 have a large mass splitting. It is even 
 possible that the lighter eigenstate 
 $\tilde t_{1}$ could be  lighter than the 
 top quark itself \cite{JEllis},
 \cite{STOP_SUSY}.

 Searches for top squarks were performed at  LEP and
 Tevatron and 
 will continue 
  at  LHC and ILC  \cite{ILCRDR1}, \cite{ILCRDR2}. 
     
 This Note is the continuation of our
 previous Note where
 we have considered stop pair production in
 photon-photon collisions \cite{BMMSS-STOP}.
  In the following we study the reaction 
\begin{equation}
      e^{+} + e^{-} \to \tilde t_{1} 
        + \bar{\tilde t_{1}} . 
\end{equation}
 
  Among the possible  $\tilde t_{1}$-decay 
 channels within
 the MSSM (see \cite{A. Bartl} for details),  
 we focus on the decay
 $\tilde t_{1} \to b \tilde \chi_{1}^{\pm}$
 followed by the two-body chargino decay 
 $\tilde \chi_{1}^{\pm} \to
 \tilde \chi_{1}^{0} W^{\pm}$,
 where one of the W's decays hadronically, 
  $W \to q_{i}{\bar q_{j}} $, 
 and the other one leptonically, $W \to \mu\nu_{\mu}$ 
 \footnote{   The process $e^{+}e^{-} \to  \tilde t_{1} +
           \bar{\tilde t_{1}}$
            with  the subsequent decay
            $\tilde t_{1} \to  c \tilde \chi_{1}^{0}$
	    was  considered
         in \cite{H.Nowak1} --\cite{H.Nowak3}.}
         \cite{Paris2004}.
  The final state of this signal process, 
  shown in the left  diagram of
  Fig.1, contains two 
 $b$- quarks, two 
 quarks  (originating  from the
 decay of  one W boson), a hard muon  plus a  neutrino 
 (from the decay of the other W) and two neutralinos:
 \begin{equation}
   e^{+}e^{-} \to \tilde t_{1} \bar{\tilde t_{1}} \to
  b\bar{b}\tilde\chi^{+}_{1}\tilde\chi^{-}_{1} \to
   b \bar{b}W^{+}W^{-}\tilde\chi^{0}_{1}
   \tilde\chi^{0}_{1} \to 
   b\bar{b}q_{i}\bar{q_{j}}\mu\nu_{\mu}\tilde
    \chi^{0}_{1}\tilde\chi^{0}_{1}. 
 \end{equation}
 The main background process is top 
 quark pair production 
 with the subsequent decay $t \to bW^{\pm}$ (for W's we 
 use the same decay channels as in the stop case): 
  \begin{equation}
    e^{+}e^{-} \to  t \bar{ t} \to 
    b \bar{b}W^{+}W^{-} \to 
    b\bar{b}q_{i}\bar{q_{j}}\mu\nu_{\mu}.
 \end{equation}
 The only difference between the final states of  stop
 and top  production (shown in the right diagram
 of Fig.1) is that in
 stop pair production there are two neutralinos 
 which are undetectable. Thus,  both processes   
 have the same signature: two $b$-jets, two jets 
 from W decay and a  muon. In the following we 
 show that the physical variables constructed of
 jets combinations may allow to  reconstruct 
  the scalar top quark mass.

 In the present paper we  consider only top pair 
 production as  background.
      \begin{figure}[!ht]
     \begin{center}
    \begin{tabular}{cc}
\mbox{a)\includegraphics[width=7.8cm, height=5.28cm]{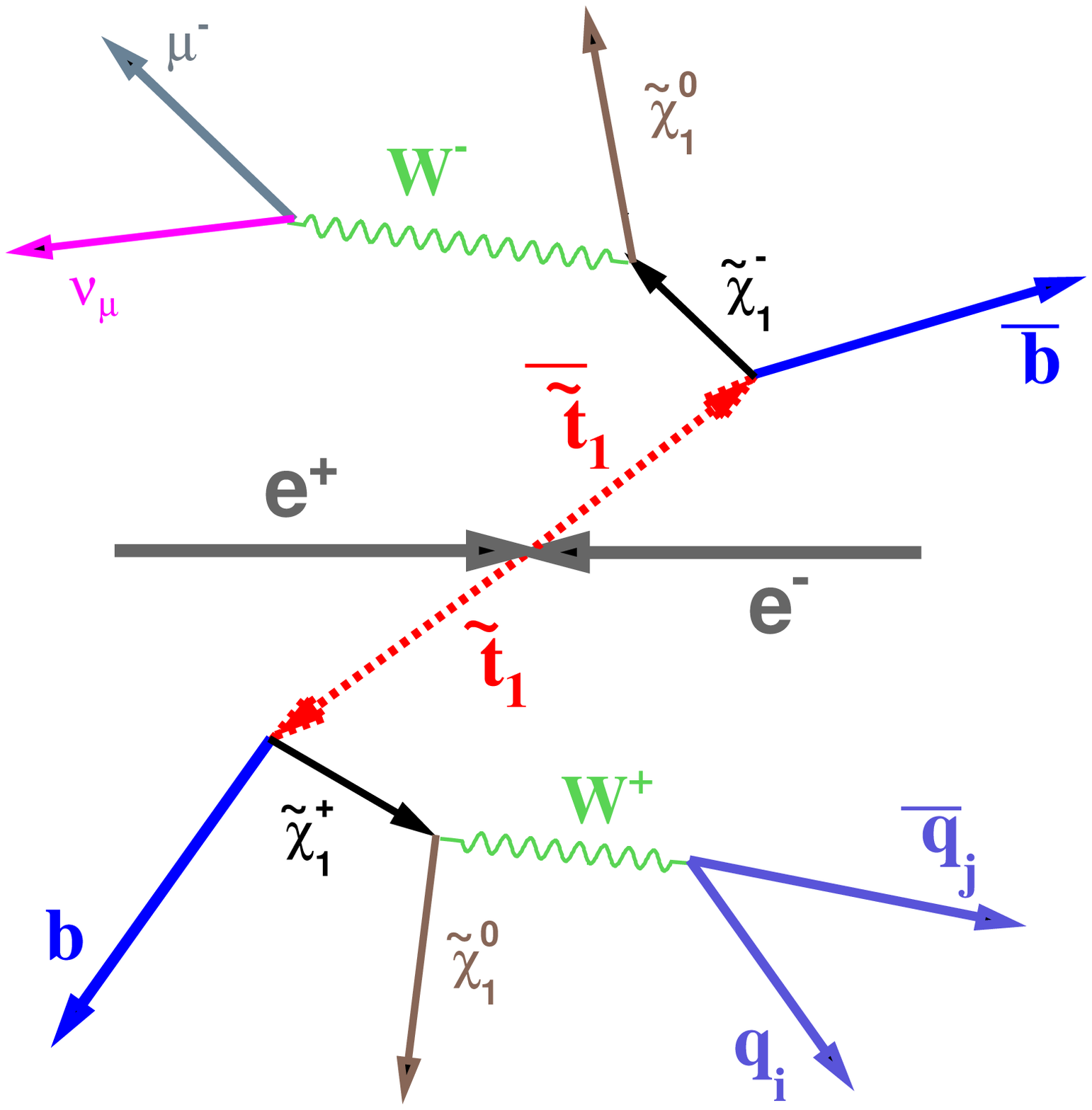}}
\mbox{b)\includegraphics[width=7.8cm, height=5.28cm]{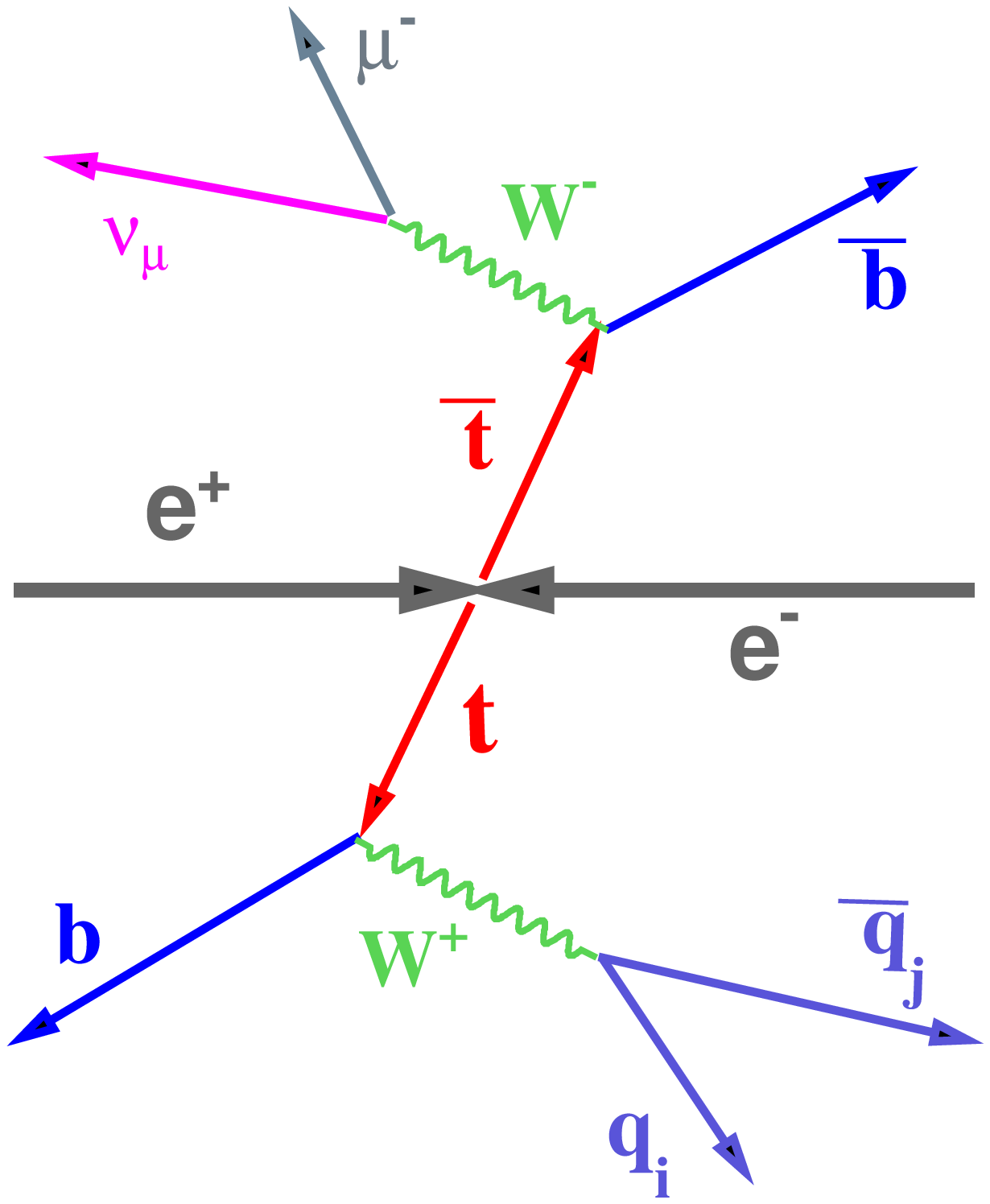}} \\
    \end{tabular}
     \caption{\it \small Left is the stop
                         signal event diagram, 
                         Right is the top 
                         background diagram. }    
     \end{center}  
  \vskip -0.5 cm           
     \end{figure}
   We analyse  the processes (2) and
  (3) with the help of  Monte Carlo
  samples of the corresponding events
  generated by the two programs PYTHIA6.4
  \cite{T. Sjostrand94}     and CIRCE1 \cite{T.Ohl}. 
   The program CIRCE1  is used for  the  
  parametrization of the beam spectra 
  involved in the processes (2) and (3) to 
  account for the effects of beamstrahlung.
    The energy of the  beams is chosen as 
  $2E_{beam} =\sqrt {s}$ = 350, 400, 500,
  800, 1000 GeV.
  Most plots are shown for 
  $\sqrt {s}$ = 500 GeV and a stop mass
  $M_{\widetilde{t}_1}=167.9$ GeV.

  In Section 2 we give the set of MSSM 
  parameters used in our  study.

  In Section 3 we discuss  some  general
  characteristics  of the signal process
  $e^{+}e^{-} \to \tilde t_{1} \tilde {\bar{t}}_{1}$
  and the main background  $ e^{+}e^{-} \to t\bar{t}$.
  First we show
how the beam energies are affected by beamstralung and other beam interaction effects
which are simulated with CIRCE1.
  The subsections include   kinematical
  distributions  (obtained without imposing 
  of any cuts) for the produced stop
  quarks,  for the jets originating from W
  boson decay and for $b$-jets.
  We compare them in detail with those of 
  top pair production.

  In Section 4 we  demonstrate  how to discriminate
  between the signal muons produced in  W 
  boson decays and those stemming from hadron 
  decays in the same events.

  In Section 5 we
  show the distributions of the 
  global variables  missing energy,
  total visible energy, the scalar sum of
  the transverse momenta of all visible
  particles in the event and the invariant
  masses of the  final-state hadronic jets
  plus the signal muon.  We show that they
  are good tools for separating the 
  signal from the top background.

   In Section 6 we  introduce further two
   global variables, 
   the invariant mass of the
   final-state hadronic jets
   and the missing mass and demonstrate  
  that they are very useful for the
   separation of signal  stop
  events from the  background 
  top events. We 
 propose  three cuts which provide
 a good  signal-to-background  ratio 
 (S/B).
      The impact of the proposed  cuts on 
  the values of the cross sections of stop and
  top pair productions are shown together with
  the values of    S/B ratios  for different
  values of $\sqrt{s}$.

  Section 7  is devoted to the
   mass reconstruction of the scalar top   quark
  based on the distribution of the  invariant mass  
  of one $b$-jet and    the other two $non-b$-jets  (from W decay), 
  provided that  the neutralino mass is known.

    In Section 8 we show 
 the distributions  of the invariant 
    variables described in Section 7 for
  a stop mass $M_{\widetilde t_1} = 200$ GeV.
   
  Section 9 contains some  conclusions.

\section{ MSSM parameters and cross section.}

  ~~~~ The scalar top quark system is described by the mass matrix 
   (in the $\tilde t_{L} - \tilde {t}_{R}$ basis) \cite{JEllis},
   \cite{Gunion}
\begin{equation}    
   \left(\begin{array}{cc} M^2_{\tilde t_{LL}}
     & M^2_{\tilde t_{LR}} \\
    
           M^2_{\tilde t_{RL}} & M^2_{\tilde t_{RR}}
	    \end{array}\right) 
\end{equation}	   
with
\begin{equation}  
 M^2_{\tilde t_{LL}} = M^2_{\tilde Q} + (\frac{1}{2} - \frac{2}{3} sin^2
 \Theta_W) cos2 \beta M^2_Z + M^2_t, 
\end{equation}
\begin{equation} 
M^2_{\tilde t_{RR}} = M^2_{\tilde U} +  \frac{2}{3} 
sin^2 \Theta_W cos2 \beta M^2_Z + M^2_t, 
\end{equation}
\begin{equation}  
M^2_{\tilde t_{RL}} = (M^2_{\tilde t_{LR}})^* =  M_t(A_t - \mu ^* cot \beta). 
\end{equation}
   The mass eigenvalues are given by
\begin{equation} 
 M^2_{\tilde t_{1,2}} = \frac{1}{2}\left[ (M^2_{\tilde t_{LL}} + 
  M^2_{\tilde t_{RR}}) \mp \sqrt{(M^2_{\tilde t_{LL}} + 
  M^2_{\tilde t_{RR}}) + 4 |M^2_{\tilde t_{LR}}|} \right]
\end{equation}
with the mixing angle
\begin{equation}
cos \theta_{\tilde t} = \frac {-M^2_{\tilde t_{LR}}} {\sqrt {
|M^2_{\tilde t_{LR}}|^2 + (M^2_{\tilde t_{1}} -  M^2_{\tilde t_{LL}} )^2}}
\end{equation}
\begin{equation}
sin \theta_{\tilde t} = \frac {M^2_{\tilde t_{LL}} - M^2_{\tilde t_{1}} } {\sqrt {
|M^2_{\tilde t_{LR}}|^2 + (M^2_{\tilde t_{1}} -  M^2_{\tilde t_{LL}} )^2}}
\end{equation}

   In the following we will consider a particular 
  choice of the MSSM parameters that are defined, in the 
  notations of PYTHIA6.4,  in the following  way: \\
   
~~  $ M_{\widetilde{Q}} = 270$ GeV;
    ~~$ M_{\widetilde{U}} = 270$ GeV;  
   ~~$ A_t = -500$ GeV (stop trilinear coupling);  
    
~~~~~~~~~~~~ $ tan \beta = 5 $;~~ $\mu =  -370 $ GeV;  
  ~~ $ M_{1} = 80 $ GeV; 
  ~~ $ M_{2} = 160 $ GeV.\\    

  Note that in PYTHIA6.4 $M_{\widetilde{Q}} $  corresponds
  to $ M_{\widetilde{t}_L} $
  (left squark mass for the third generation) and 
  $M_{\widetilde{U}}$  corresponds to $M_{\widetilde{t}_R}$.
  These parameters give $M_{\widetilde{t}_1}=167.9$ GeV, 
  $M_{\chi^{+}_{1}}=159.2$ GeV 
  and  $M_{\chi^{0}_{1}}=80.9$ GeV. 
  This parameter point is compatible
  with all experimental
  data. We have chosen
  this value of $ M_{\widetilde{t}_1}$
  rather close
  to the mass of the top quark 
  $M_{top}=170.9 \pm 1.8$ GeV 
  \cite{Schieferdecker}.
  This means that one expects a rather 
  large  contribution 
  from the top background,
  therefore, this choice 
 makes the analysis 
  most difficult. Finding a suitable set of 
  cuts separating stop and top events 
  is crucial.
 
   In general, the cross section for stop pair 
   production at a
   fixed energy depends on the mass of the 
   stop quark and the mixing angle
   $\theta_{\tilde t}$. Since the couplings
   of the $Z^0$ to the left and right
   components of the stop are different,
   the cross sections depend significantly
   on the beam polarizations 
   (see \cite{A. Bartl}, \cite{Paris2004},
   \cite{Gudi}). By choosing 
   appropriate
   longitudinal beam polarizations it is possible
   to enhance the cross sections. For
   example, for an electron beam with 90$\%$
   left polarization the cross section
    would be larger than the unpolarized
   cross section by approximately 40$\%$, for
    cos $\theta_{\tilde t}$ = - 0.81
   corresponding to the parameters given 
   above. If in addition the positron
   beam has 60$\%$ right polarization, then the
   cross section is enhanced by approximately
   a factor of 2 compared to the 
   unpolarized cross section. We note that
   a rather precise determination of the
   stop mixing angle $\theta_{\tilde t}$
   is possible by measuring the left-right 
   asymmetry. The cross section for
   top pair production has also a
   characteristic dependence on the beam
   polarizations \cite{Gudi}. For example, 
   the polarization of both beams
   leads to an increase of the cross 
   section by about a factor of 1.5. 

\section{Distributions of kinematical 
         variables in          stop and top production.}

 ~~~~  In the following we present some 
 distributions  of different  physical variables
  based on the  sample of
  $5 \cdot 10^{4}$ stop pair production 
  events generated by  PYTHIA6.4 and CIRCE1  
  weighted with   the electron-positron
  luminosity. Analogous plots are also given for
  $ 10^{6}$  generated  background top events. 
  In this Section all plots are obtained 
  without   any cuts.

  The ILC is a  linear electron-positron 
  collider with a center-of-mass energy of
  200-500 GeV and a high luminosity
  (peak luminosity of 
  $\sim 2\cdot 10 ^{34} cm^{-2}s^{-1}$),
  upgraded  to 1~TeV   in the second phase.  
  According to \cite {ILCRDR1}  
  a total luminosity  of 500 fb$^{-1}$ 
  is foreseen within  the first four years of 
  operation  and  1000 fb$^{-1}$ 
  during the first phase of operation
  at  500 GeV.  A first   run 
  at  $\sqrt {s}=500$ GeV   
  will get a first measurement of 
  the particle masses to optimize
  the threshold scan \cite{ILCRDR2}.
  
   Fig.2 {\bf a)} demonstrates the total    energy  
   spectrum of the electron and 
   positron beams $E1_{e^+ } +E2_{e^-}$,
   which is expected at 
   $2E_{beam}=\sqrt {s}=500$ GeV after 
   taking into account   beamstrahlung and 
   other beam interaction effects (see, for 
  instance \cite{DShulte}).
  Fig.2 {\bf b)} shows the correlations 
  of the beam energy fractions
  $y_{i}=E_{i}/E_{beam}$ (i=$e^+,e^-$)
   of the colliding 
  electron and positron  beams.    
 \begin{figure}[!ht]
     \begin{center}
    \begin{tabular}{cc}
  \mbox{a)\includegraphics[width=7.2cm, height=5.2cm]{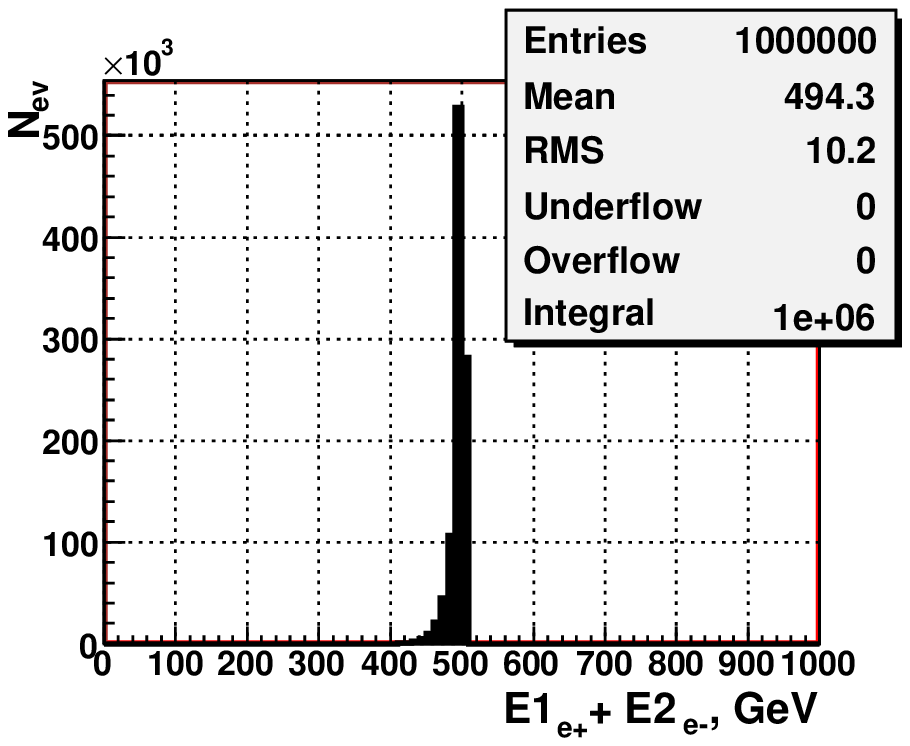}}        
 \mbox{b)\includegraphics[width=7.2cm, height=5.2cm]{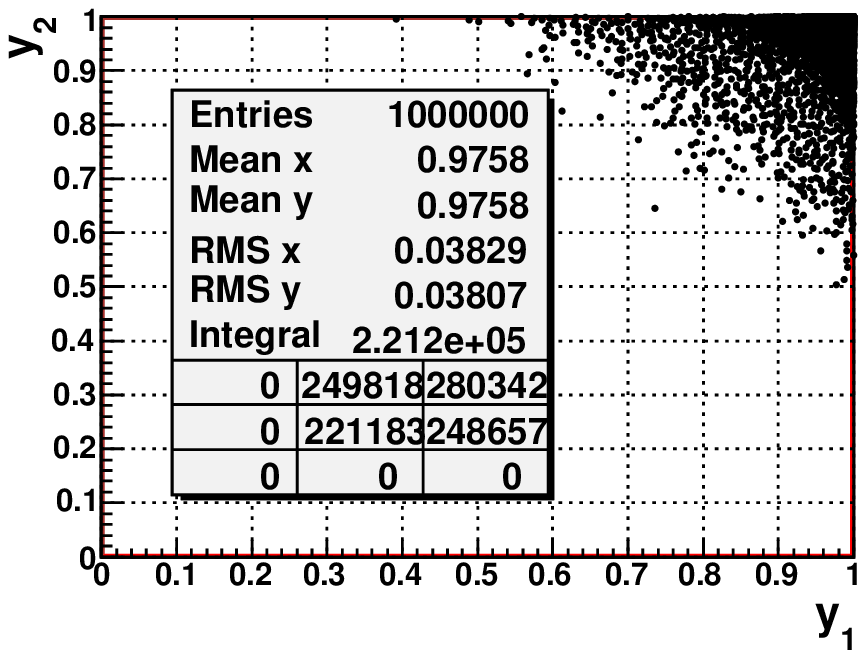}} \\
    \end{tabular}
     \caption{\small \it {\bf a)} total beam energy spectrum,
     {\bf b)} beam fractions correlations. }     
     \end{center}   
 \vskip -0.5 cm           
     \end{figure}

\subsection{ Distributions in stop events.}

~~~  In Sections 3, 4 and 5  we will present plots for the 
  kinematical distributions only for the energy 
   $\sqrt {s}=500$ GeV.
      Fig.3 shows  two  kinematical 
  distributions characteristic of the  produced 
  stop pair system, i.e., the  
  number of expected events versus the
  stop transverse  momentum  
  $PT_{\widetilde{t}_1} $  (plot {\bf a)})
  and  its polar angle 
  $\theta_{\widetilde{t}_1}$ (see plot  {\bf b)} 
   (all in the $e^{+}e^{-}$ c.m.s.).
   As can be seen  in  Fig.3 {\bf a)}  the stop transverse 
   momentum  $PT_{\widetilde{t}_1} $
   spectrum begins at
   $PT_{\widetilde{t}_1} \approx 50$ GeV
   and has a peak near the kinematical  limit.

    \begin{figure}[!ht]
    \begin{center}
    \begin{tabular}{cc}      
     \mbox{a) \includegraphics[width=7.2cm, height=5.2cm]{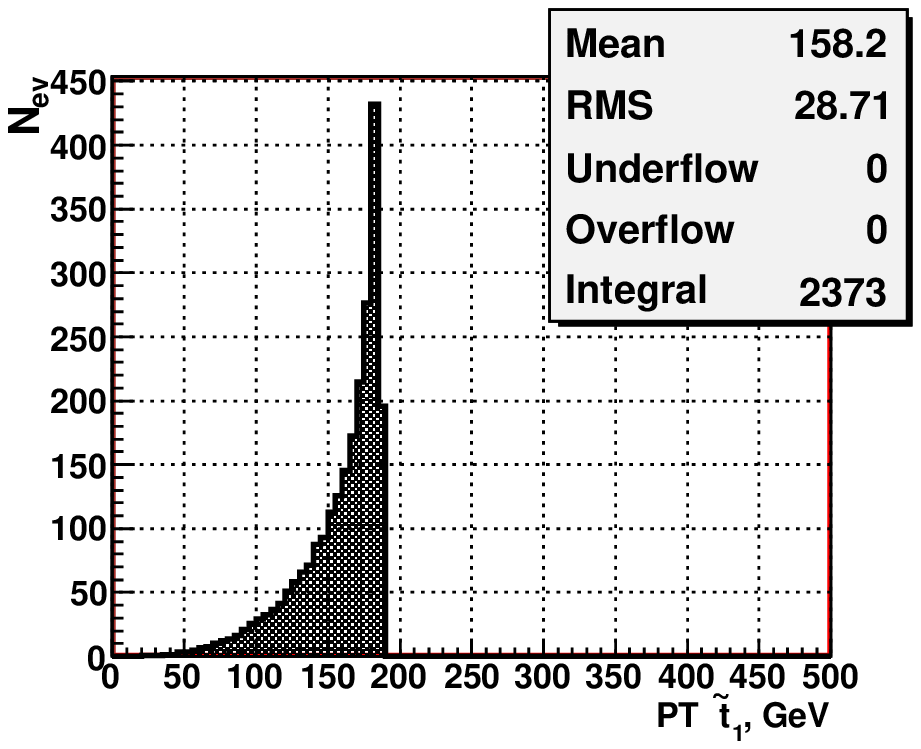}}      
     \mbox{b) \includegraphics[width=7.2cm,
      height=5.2cm]{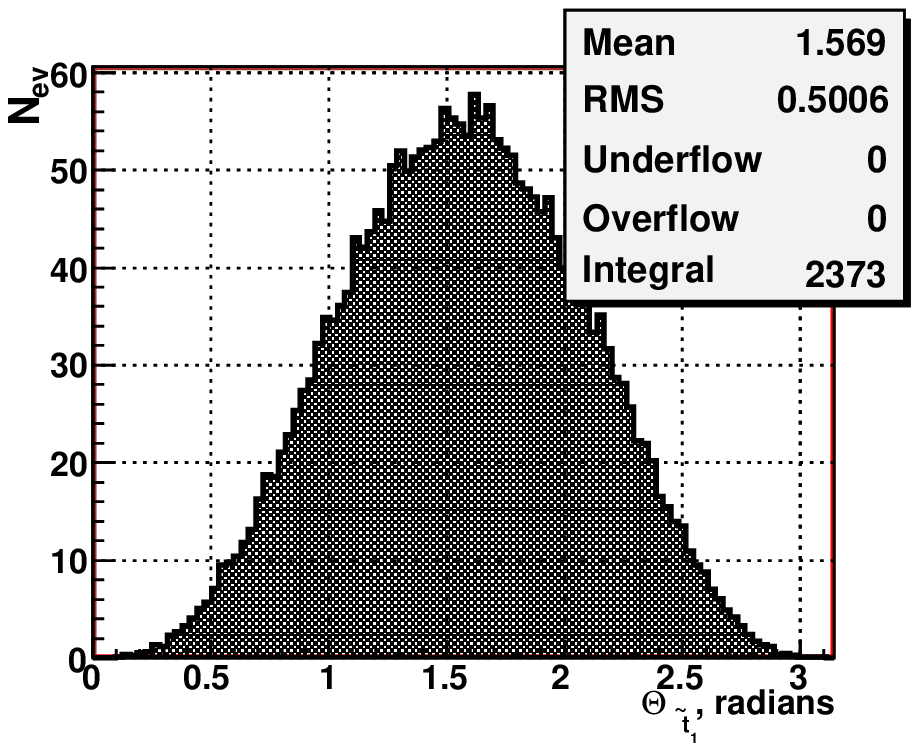}}      
   \end{tabular}
 \caption{\small \it  
       Distributions of the
        number of events
        $N_{event}$ $(L_{int}=1000 fb^{-1})$ versus:
        {\bf a)} stop transverse momentum
        $PT_{\widetilde{t}_1} $;   
        {\bf b)} stop polar angle   
       $\theta_{\widetilde{t}_1}$.}       
     \end{center}
 \vskip -0.5cm           
     \end{figure}

   In Fig.3 and the following figures the vertical axis
   shows the number    of stop and top events 
   that may be expected 
   for the integrated luminosity 
   L$_{int}=$ 1000 $fb^{-1}$.
   Taking the integral of the distributions 
   one can get  the total  number  of events expected
   for this  luminosity.
   These numbers   are shown as  "Integral" values 
   in the Figures. 
   
  To find the jets we  use
  the subroutine PYCLUS of PYTHIA  with the
  distance measure  utilized
  in the "Durham algorithm". The parameters 
  of this  jet finder,
   which is widely used in $ e^{+}e^{-}$
  physics,  are  chosen such that 
  the number of jets is $exactly$ four (see
  also \cite{MoenigKlamke}).

%
\subsection{  Distributions of the jets from  W decay.} 
%

 ~~~~  According to the decay chain (2), the final 
   state has to contain two jets
   due to the decay of one W boson  into 
   two quarks $W \to q_{i} + \bar q_{j}$
   (see Fig.1).
 
    \begin{figure}[!ht]
     \begin{center}
    \begin{tabular}{cc}
  \mbox{a) \includegraphics[ width=7.2cm, height=5.2cm]{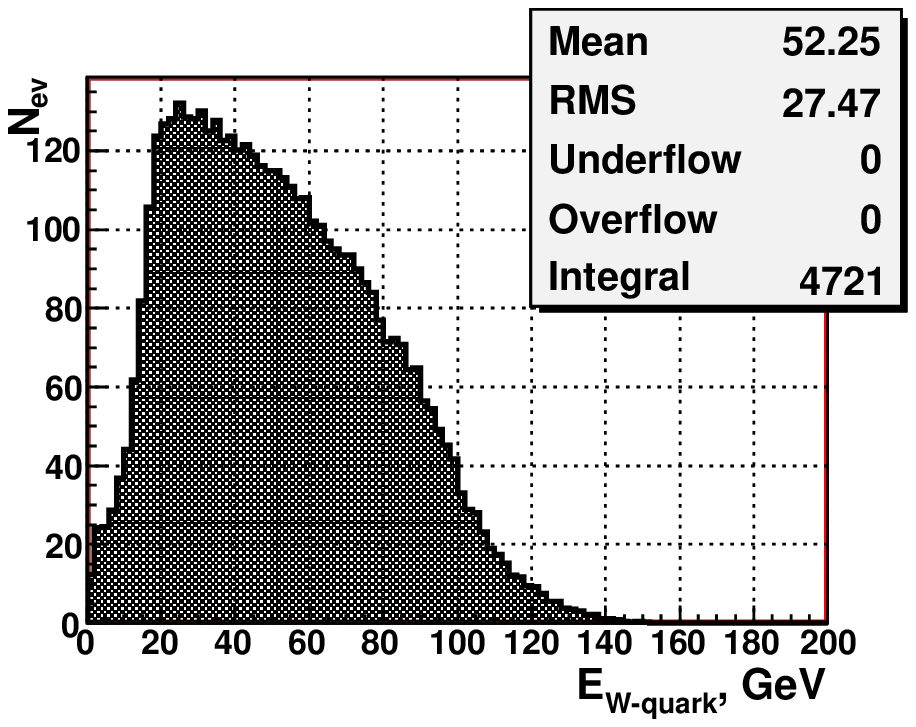}}
  \mbox{b) \includegraphics[ width=7.2cm, height=5.2cm]
{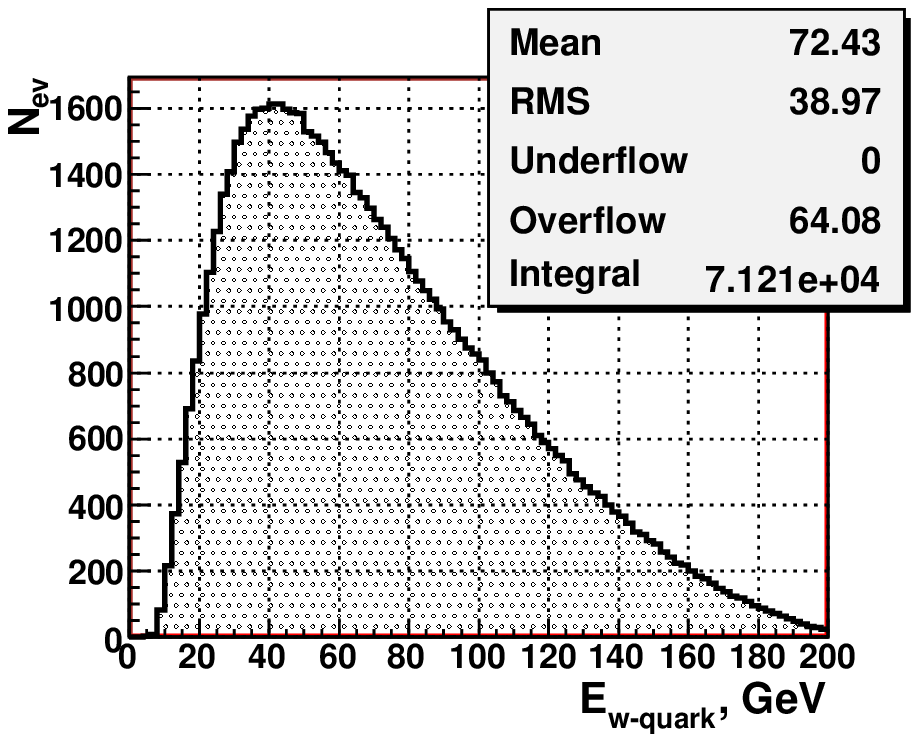}} 
    \end{tabular}
     \caption{\small \it Energy spectra of 
                the quarks from W boson decay.  
      	  {\bf a)} stop pair production;
	  {\bf b)} top pair production.}
     \end{center} 
\vskip -0.5cm            
     \end{figure}

     Plot {\bf a)}  (for stop pair production) 
   and  plot  {\bf b)}    (for top pair production)
   of Fig.4 show the distributions of the energy
   $E_{W-quark}$ 
   of the quarks  produced in the 
   W boson decay (which we  call "$W$-quarks").  
   The first spectrum begins  at zero and  
   goes up to 140 GeV, with a mean value of 52 GeV,
   while the second spectrum begins
   around 8 GeV and goes 
   up to approximately 200 GeV,
   with a mean value of 72 GeV.

    \begin{figure}[!ht]
     \begin{center}
    \begin{tabular}{cc}
        \mbox{a) \includegraphics[  width=7.2cm, height=5.2cm]{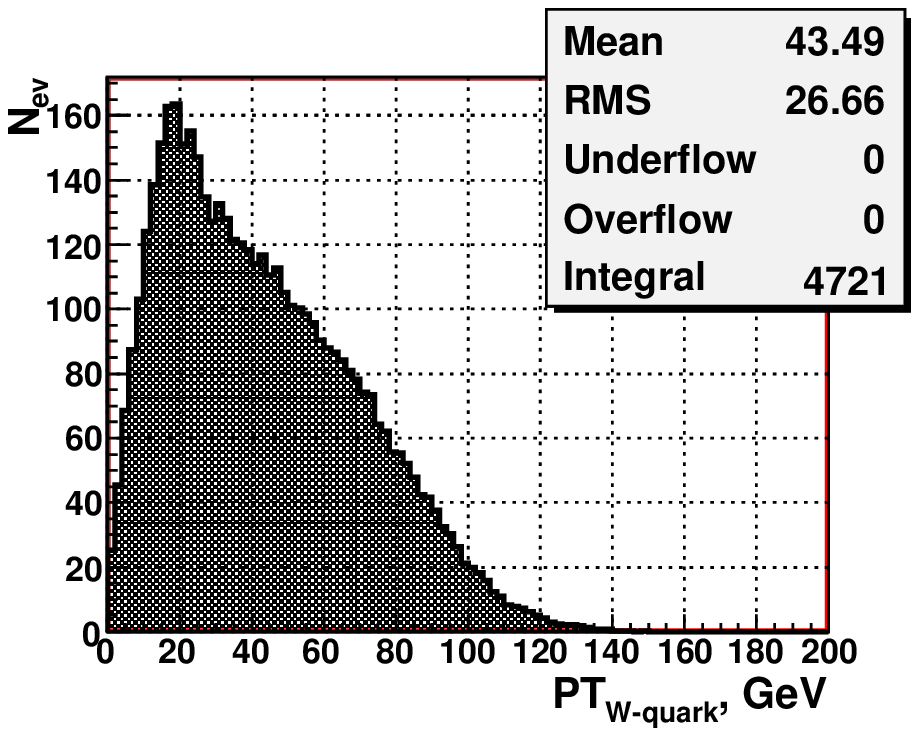}}      
        \mbox{b) \includegraphics[  width=7.2cm, height=5.2cm]{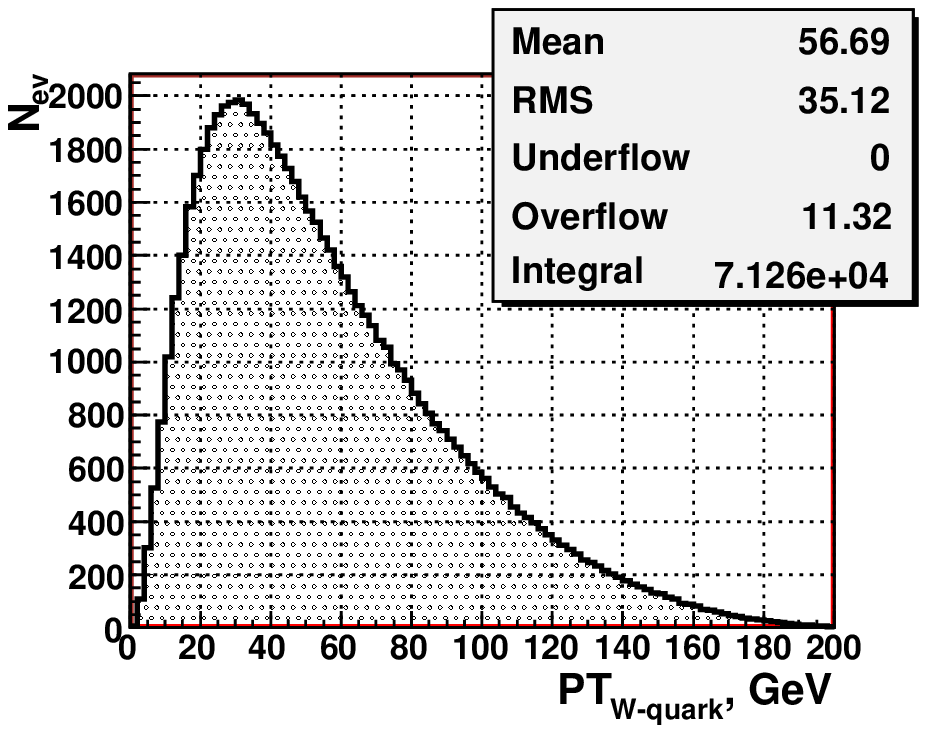}} 
     \end{tabular}
     \caption{\small \it PT spectra of the quarks   
                     produced in the W boson decay.
           {\bf a)} stop pair  production;
	   {\bf b)} top pair production.}
     \end{center}  
\vskip -0.5cm             
     \end{figure} 
      
   Figure 5 shows the transverse momentum 
   $PT_{W-quark}$ spectra  of the quarks produced 
   in the W boson decay  for stop (plot {\bf a)})
   and  top (plot {\bf b)}) production. 
   The shapes of the  $PT_{W-quark}$ spectra of 
   these "$W$-quarks"  are  rather 
   similar  to the
   $E_{W-quark}$ spectra. In the case of top 
   production the "$W$-quarks" are slightly 
   more energetic and have a larger transverse 
   momentum than those from  stop pair production.

   As the next step we take into account the
   hadronization of the "$W$-quark" into a jet which we call "$jet_W$".
    Figs.6 and 7 show the energy
   $E_{jet_W}$  and  transverse momentum  $PT_{jet_W}$
   distributions  of the corresponding "$W$-jets".
   Plots  {\bf a)} and {\bf b)}  are for  stop
   and  top production, respectively.
   According to our  
   choice  of PYCLUS jet finder 
   parameters there are   two
      "$jet_W$"  in the event.
   
     \begin{figure}[!ht]
     \begin{center}
    \begin{tabular}{cc}
 \mbox{a) \includegraphics[   width=7.2cm, height=5.2cm]{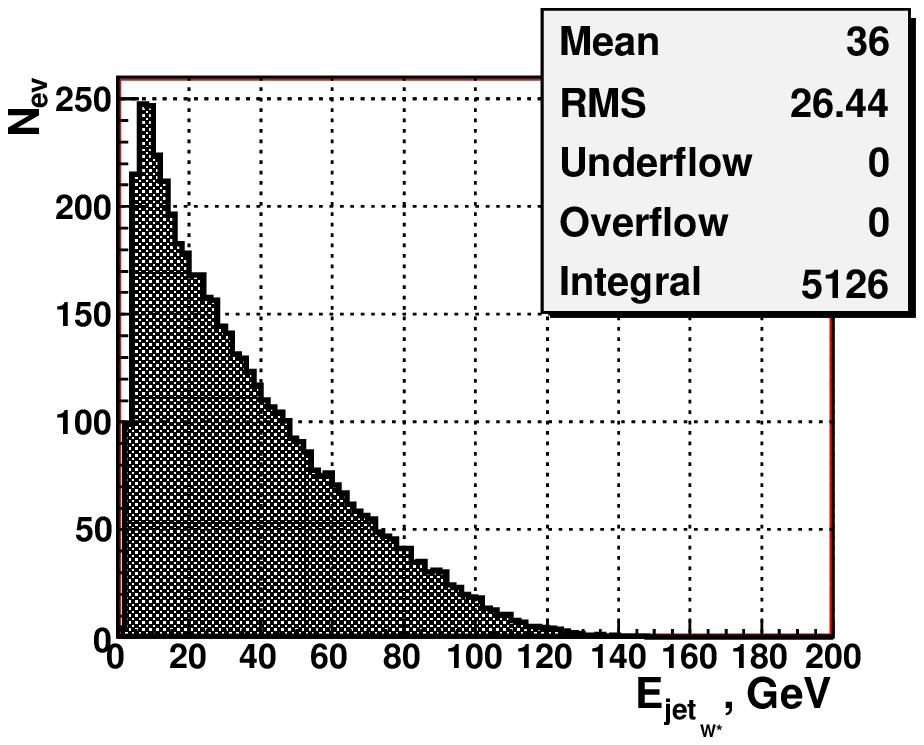}}
 \mbox{b) \includegraphics[  width=7.2cm, height=5.2cm]{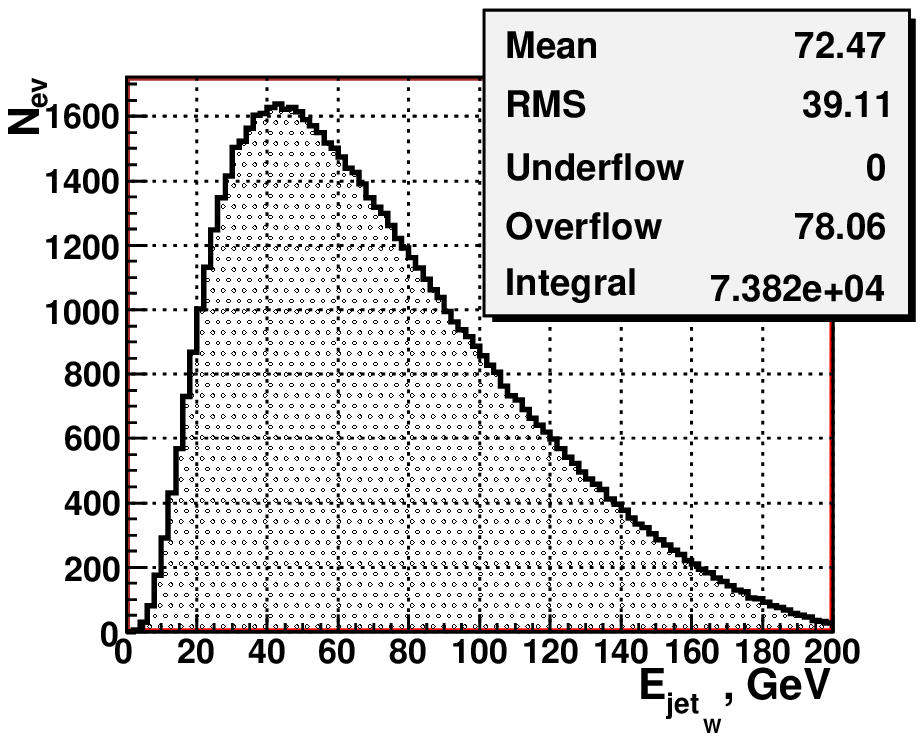}} 
     \end{tabular}
     \caption{\small \it $E_{jet_W}$ energy spectra.
             {\bf a)} stop pair production;
	     {\bf b)} top pair production.}
    \end{center} 
\vskip -0.5cm              
     \end{figure}

   Comparing plot {\bf a)} of Fig.6 
   for the  energy distribution  of "W-jets" 
  in  stop production with     the  plot {\bf a)} of Fig.4  
   one observes that 
   the corresponding mean
   value of the  "W-jets" 
   energy  $E_{jet_W{*}}$ 
   (we use the notation $W^{*}$ in the stop case, for
   some details see below) in  Fig.6
   is about 16 GeV lower than  the
   mean  energy 
   $E_{W-quark}$  of "$W$-quarks".
   It is also seen (plot {\bf a)} of Fig.4)
   that the peak position of 
   "$W$-quark" energy distribution 
   ($E^{peak}_{W-quark} \approx 25$ GeV) 
   is shifted to the left by about 17 GeV
   ($E^{peak}_{jet_W{*}}\approx 8$ GeV)
   when passing to the jet level
   (see plot {\bf a)} of Fig.6).
   The end point of the $E_{jet_W{*}}$ 
   distribution in stop case is somewhat lower
   than that one for the corresponding  quarks. 

     \begin{figure}[!ht]    
    \begin{center}
    \begin{tabular}{cc}
   \mbox{a) \includegraphics[width=7.2cm, height=5.2cm]{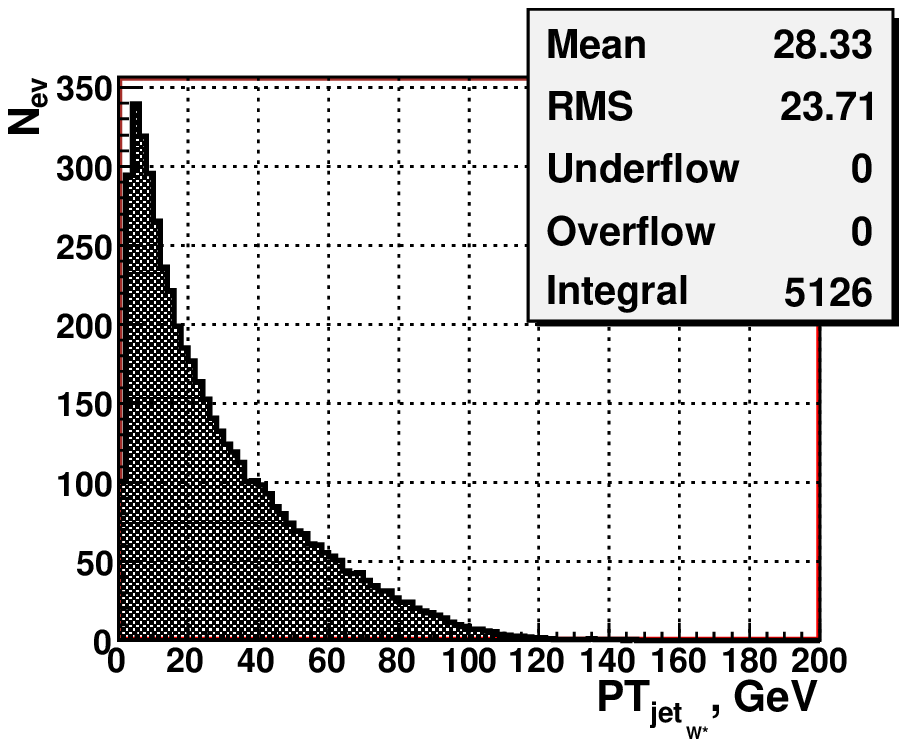}}
    \mbox{b) \includegraphics[width=7.2cm, height=5.2cm]{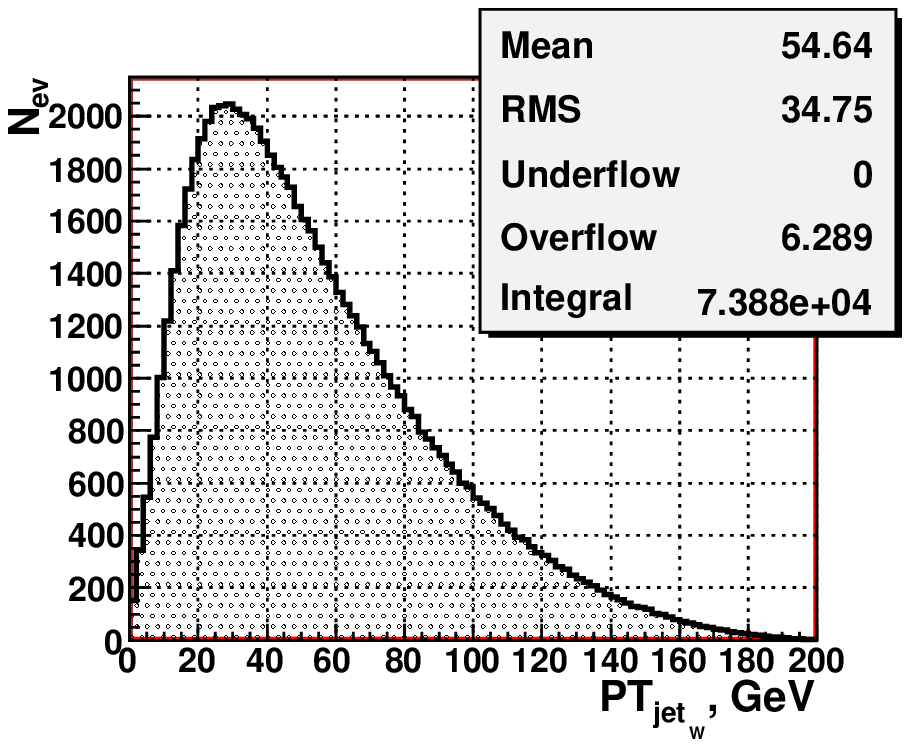}} 
     \end{tabular}
     \caption{\small \it   "$jet_W$"  PT- spectra.  
               {\bf a)} stop pair production;
	       {\bf b)} top  pair production.}                       
     \end{center} 
\vskip -0.5cm              
     \end{figure}

   Analogously, the mean value and the peak 
   position of the distribution of the  transverse
   momentum of the "$W^{*}$-quarks", $PT_{W^{*}-quark}$, 
   shown in Fig.5 {\bf a)},
   decrease by about 13-15 GeV when passing 
   to the jet level (see Fig.7 {\bf a)}),
   while the  end point of the
   $PT_{jet_{W}{*}}$ distribution is a bit lower than 
   the end point of 
   $PT_{W^{*}-quark}$ distribution.
   
   Due to the  different 
   kinematics in  top production mentioned above,
   the energy $E_{jet_W}$,   its peak position 
   and the mean value of the   "$jet_W$"
   energy distribution in the top case 
   are practically equivalent  to the
   $E_{W-quark}$  spectrum,    peak position and the
   mean value of the 
   corresponding  "$W$-quark" energy
   distribution (see Fig.4 {\bf b)} 
   and Fig.6 {\bf b)}).
Analogously, by comparing plots {\bf b)} of Figs. 5 and 7 for $PT_{W-quark}$ and $PT_{jet_W}$, one can see that the transverse momentum distribution in top production is stable under hadronization.

 \begin{figure}[!ht]
     \begin{center}
    \begin{tabular}{cc}
     \mbox{a) \includegraphics[width=7.2cm, height=5.2cm]{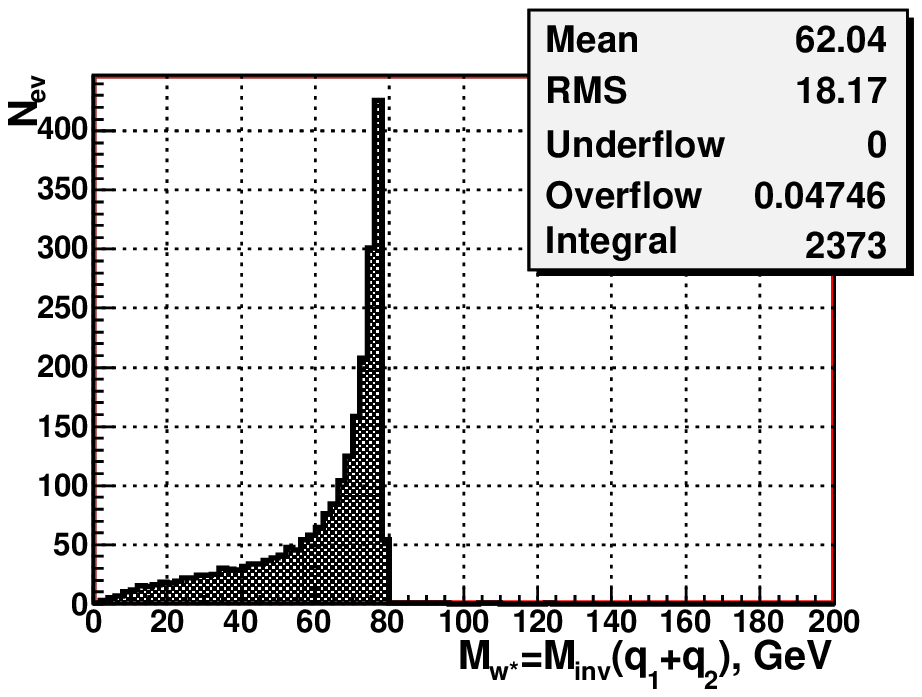}}      
     \mbox{b) \includegraphics[ width=7.2cm, height=5.2cm]{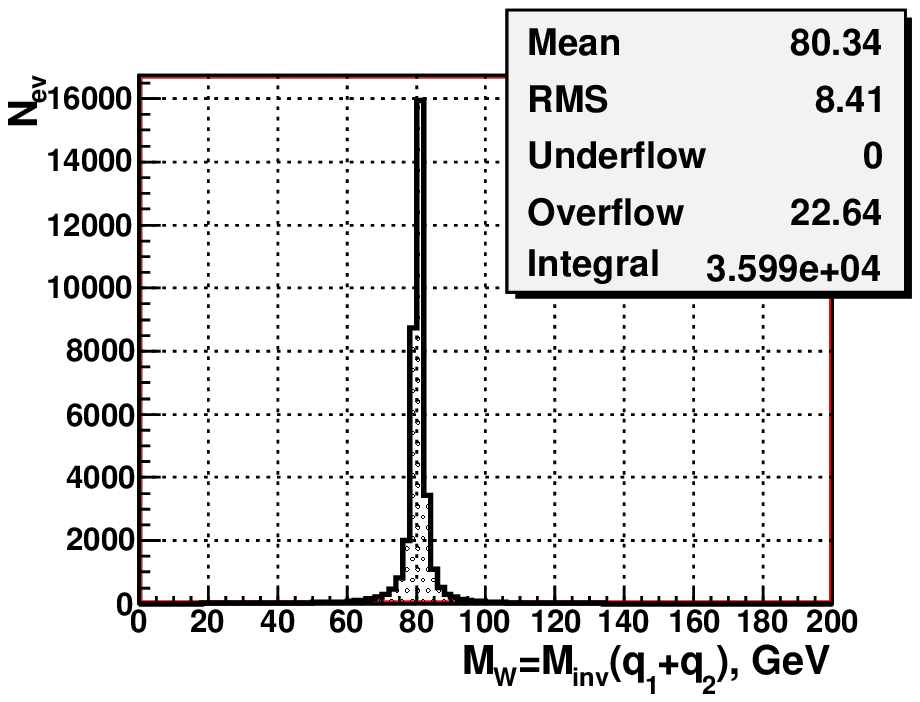}}
    \end{tabular}
     \caption{\small \it The invariant mass of two 
          quarks $M_{W}=M_{inv}(quark1 + quark2)$, 
	      reconstructed from the  vectorial
	      sum of 4-momenta of two quarks that
      are produced in  $W \to q_{i} + \bar q_{j}$ 
      decay. {\bf a)}  stop pair production;
	     {\bf b)}  top pair production.}     
     \end{center}   
 \vskip -0.5 cm          
     \end{figure}
    
      Figure 8 shows  the spectrum   of the 
  invariant mass  
    $M_{inv}(quark_{i}  + quark_{j}) \equiv M_{W} $
   reconstructed from the vectorial sum
   of 4-momenta of the two quarks produced
   in W decay $W \to q_{i}\bar q_{j}$.  
   Plot {\bf a)} is for  stop pair production, 
   plot {\bf b)} is for  top pair  production.
    In plot {\bf a)} of Fig.8 one clearly 
    sees the virtual nature of the W boson  
    in the stop pair  production case
    which we  denote by a (*) in 
    $W^{*}$.
    Hence, in the stop case the invariant mass 
    of two quarks  produced in the decay of the 
    virtual $W^{*}$ is smaller than the 
    mass of a real W boson. In  top production 
    (see plot {\bf b)} of Fig.8) there is a 
    peak in the invariant mass  distribution 
    at the mass value of the real  W boson.
  
      \begin{figure}[!ht]
     \begin{center}
    \begin{tabular}{cc}
     \mbox{a) \includegraphics[ width=7.2cm, height=5.2cm]{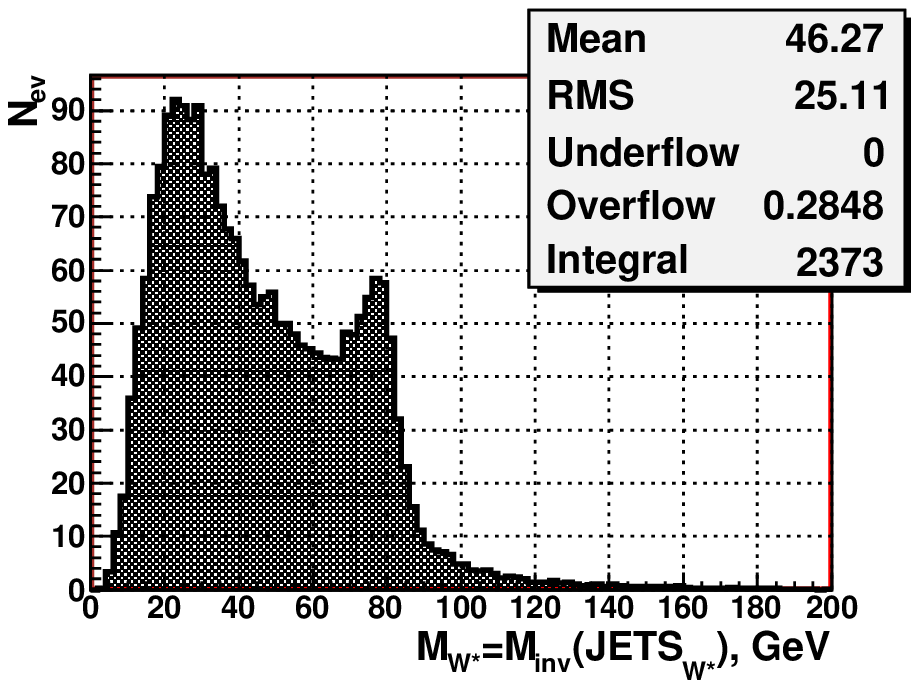}}      
     \mbox{b) \includegraphics[width=7.2cm, height=5.2cm]{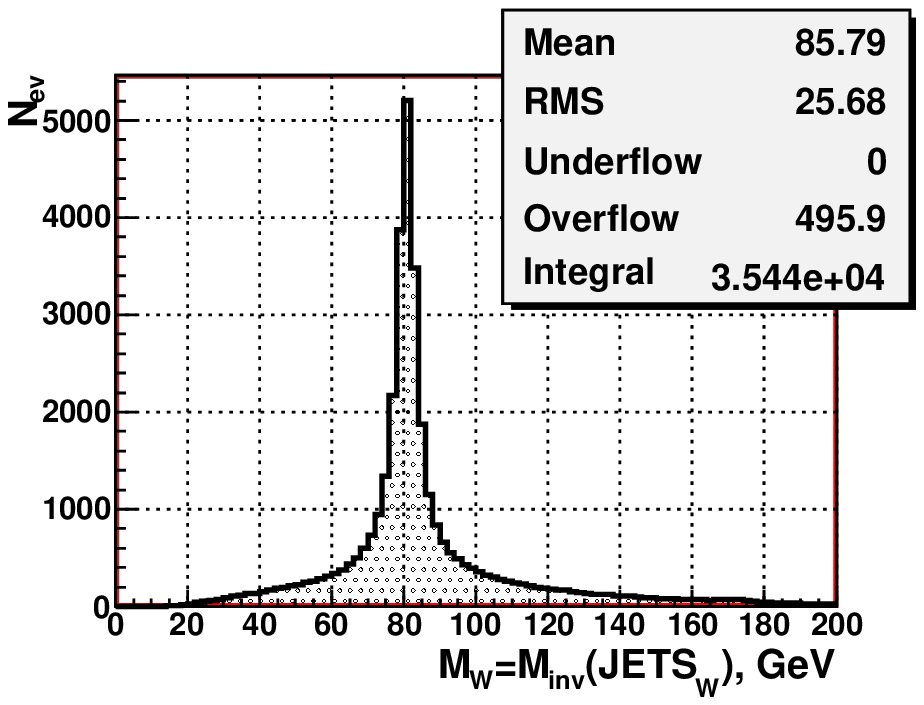}} \\
    \end{tabular}
     \caption{\small \it Number of generated events
              versus the reconstructed invariant
	      mass of "all-non-$b$-jets". 
	    {\bf a)}  stop pair production; 
	    {\bf b)} top pair production.}     
     \end{center}   
 \vskip -0.5 cm           
     \end{figure}

     Figure 9 shows  the corresponding  plots 
    at the jet level. The invariant mass in  
    Fig.9  is built of 
    two "$jet_W$" - jets  produced in the W
    boson decay    (or, shortly,  ``$JETS_{W}$'').
    One can see from plot {\bf a)} that in 
    the stop case the   spectrum 
    of the invariant masses
    $M_{W^{*}} \equiv M_{inv}(JETS_{W^{*}})$    
    is strongly shifted
    to the left. It has the main peak in the
    interval from 20 to 30 GeV and two tails.
    The left tail is rather short while the 
    right one is very long and it spreads up 
    to $M_{W^{*}} \approx 180$ GeV. There is 
    another not so high peak of, approximately, 
    $65\%$ of the height of the  main peak, 
    which is seen near the position of the peak
    shown in the quark level plot {\bf a)} of
    Fig.8  near the point  $M_{W^{*}}=80$ GeV. 
    As seen from plot {\bf b)}, in the 
    top case the position of the
    W-peak  at the jet level
    remains (with a high precision) 
    at the same value of $M_{W}$ as
    at the quark level
    in plot {\bf b)} of Fig.8. There 
    appear a shift in the mean value
     (about 6 GeV) and some     tails on  both sides.
  
%
\subsection{ $b$-quark and $b$-jet distributions in  stop 
             and top production.} 
%
   ~~~ In the case of stop  decay into a  
   $b$-quark and a chargino,
   $\tilde t_{1} \to b \tilde \chi_{1}^{\pm}$, 
   the  jets produced in $b$-quark hadronization
   are  observable objects. Their features are 
   interesting from the viewpoint of 
   experimentally distinguishing 
   the stop signal events from 
   the top background.
  
    \begin{figure}[!ht]
     \begin{center}
    \begin{tabular}{cc}
     \mbox{a) \includegraphics[  width=7.2cm,
      height=5.2cm]{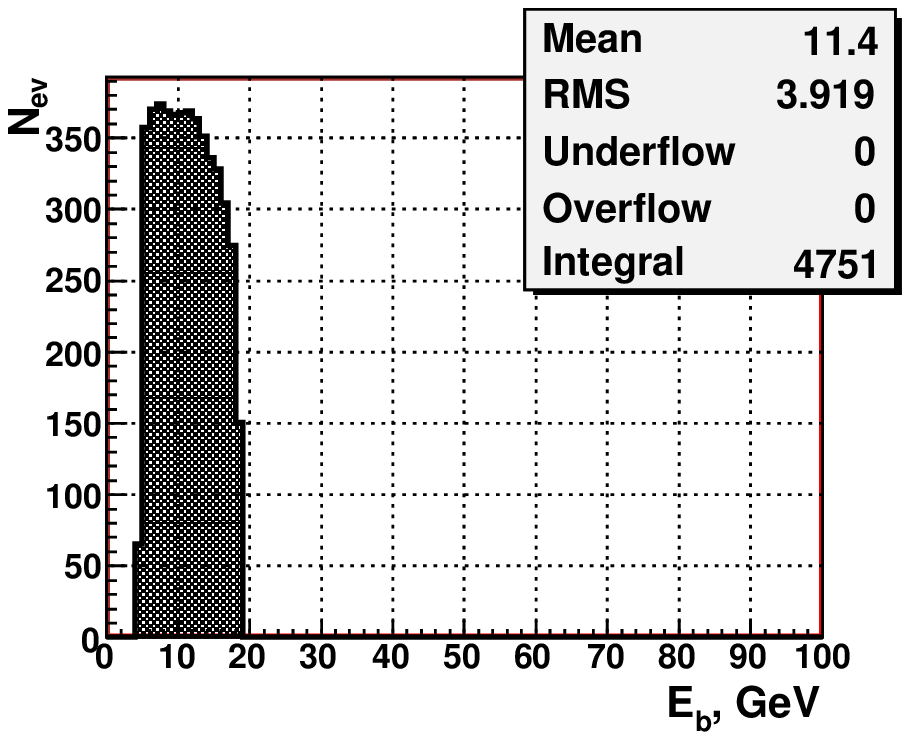}}      
     \mbox{b) \includegraphics[  width=7.2cm,
      height=5.2cm]{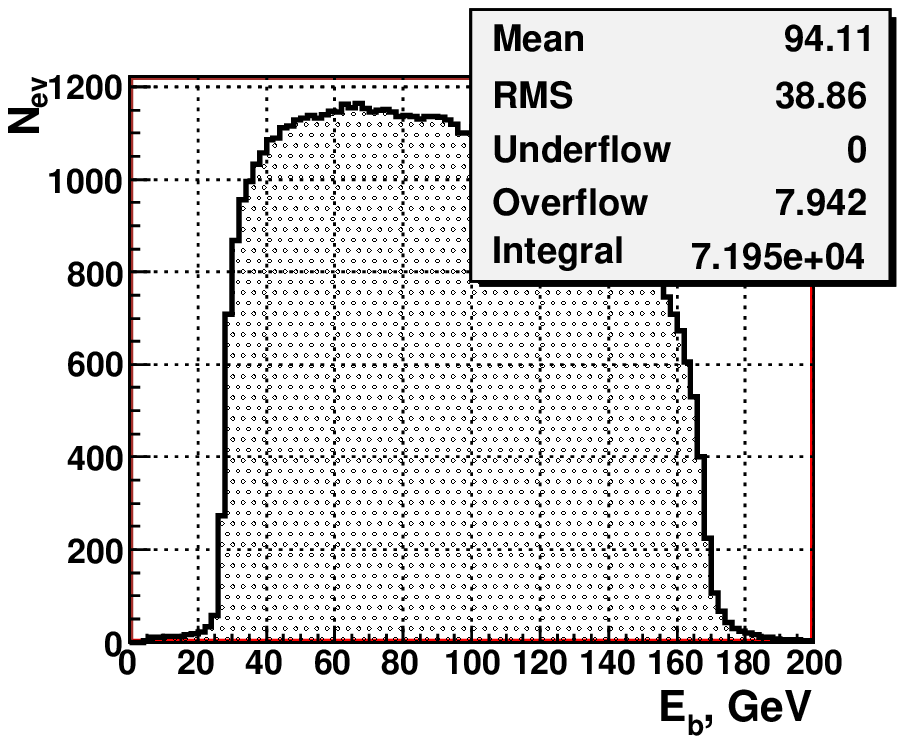}} 
    \end{tabular}
     \caption{\small \it $b$- and $\bar b$-quark 
              energy spectra.
	  {\bf a)} stop pair production;
	  {\bf b)} top pair production.}
     \end{center} 
\vskip -0.5cm            
     \end{figure}

In Fig.10 we show in     plot {\bf a)} for stop pair production 
   and plot  {\bf b)} for top pair production  
   the distributions of the energies $E_b$ of 
   the  $b$- and $\bar b$-quarks (which we do not 
   distinguish  in the following) produced 
   in  stop and top decay chains (2) and
   (3), respectively. 
   Both  spectra begin at $E_{b} \approx 4$ GeV, 
   corresponding to the b-quark mass, but
   look very different.  The b-quark
   energy spectrum in stop production strongly increases
    up at $E_{b} \approx 4$ GeV 
   and strongly decreases at 
   $E_b \approx$ 17-19 GeV. The corresponding 
   spectrum in top production is much harder
   and its main part is concentrated within
   the interval $25 < E_b  < 170$ GeV. 
   The mean  values of  the b-quark energies 
   are about 11 GeV and 94 GeV in stop and top
   production,  respectively.    
   This means that in the stop case (see plot 
   {\bf a)}) the b-quark takes a smaller part
   of the stop energy $E_{\widetilde{t}_1} 
   \approx 250$ GeV  than the b-quark
   gets in the  background top case 
   (see plot {\bf b)}).

    Figure 11 shows    the transverse momentum 
   $PT_{b}$ spectra of $b$-quarks for stop
   (plot {\bf a)}) and top (plot {\bf b)})
   production. Comparing plot {\bf a)} of 
   Fig.11 with plot  {\bf a)} in Fig.3, one 
   can conclude  that in stop 
   pair production  the $b$-quarks have
    only a small fraction 
   of the transverse momentum of the parent 
   stops. The shape of the  $PT_{b}$   spectrum 
   of $b$-quarks  in the stop case (see plot
   {\bf a)} in Fig.11) is  similar to 
   the shape of the $E_{b}$ spectrum
   (see plot {\bf a)} in Fig.10). 
   This means that in the
   stop decay the  transverse component of 
   the $b$-quark momentum is larger than
   the longitudinal component. 

    \begin{figure}[!ht]
     \begin{center}
    \begin{tabular}{cc}
     \mbox{a) \includegraphics[width=7.2cm,
      height=5.2cm]{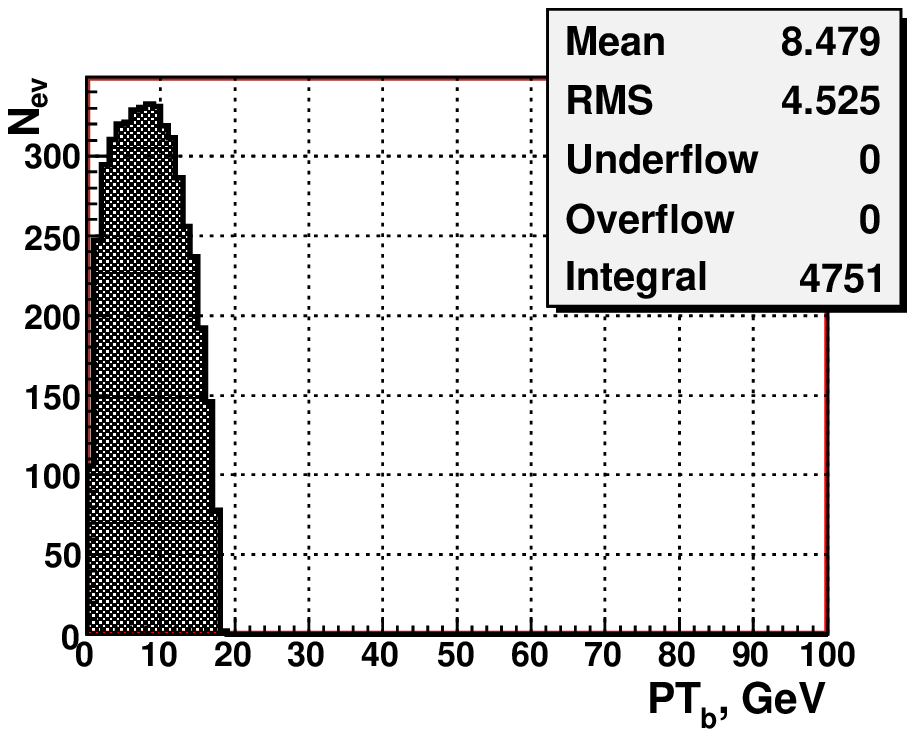}}  
         
     \mbox{b) \includegraphics[ width=7.2cm,
      height=5.2cm]{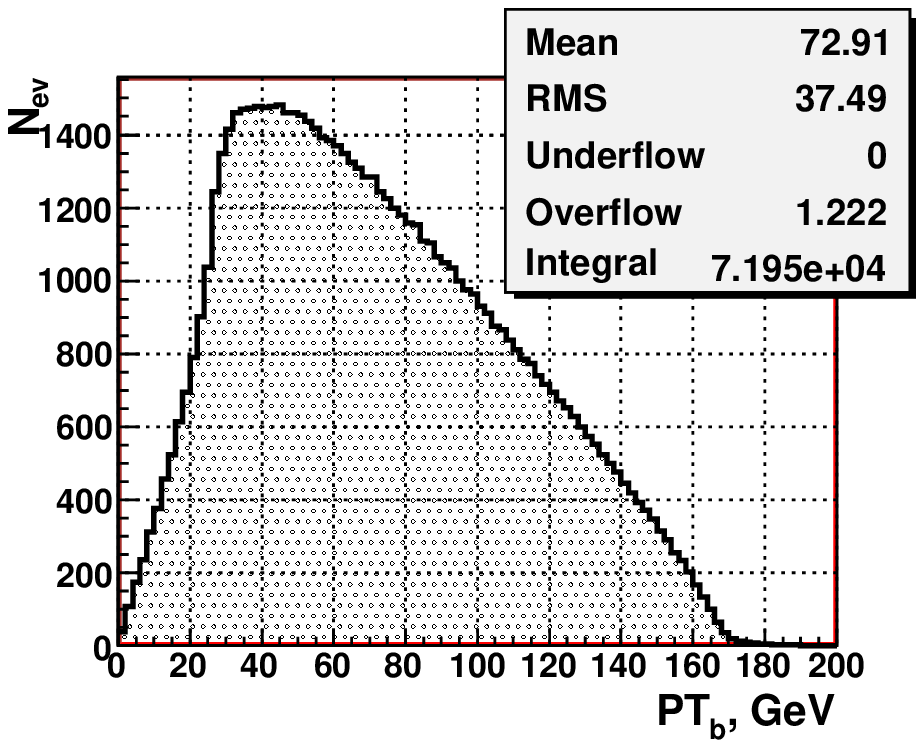}} \\
     
    \end{tabular}
     \caption{\small \it  b- and $\bar b$-quark PT spectra.
           {\bf a)} stop pair  production;
	   {\bf b)} top pair production.}
     \end{center}  
\vskip -0.5cm             
     \end{figure} 
        
   The kinematical distributions 
   of the $b$-quarks in  top decay  are quite 
   different. As seen from plots {\bf b)} of 
   Fig.10 and Fig.11,  the $b$-quarks produced in 
   the  top decays are very energetic. 
   Most of the top events have 
   $E_{b} \geq 25$  GeV and 
  $PT_{b} \geq 20$ GeV. The difference to
   stop decay is easily understandable. 
   The stop decays into a heavy chargino,
   whereas the top decays  into a real
   W boson whose mass is only half of the mass
   of the chargino  $M_{\chi_{1}^\pm}$. 
   Therefore, the  $b$-quarks  in  
   top decays have a larger phase space 
   than the $b$-quarks  in  
   stop decays.
 
     \begin{figure}[!ht]
     \begin{center}
    \begin{tabular}{cc}
  \mbox{a) \includegraphics[  width=7.2cm, height=5.2cm]{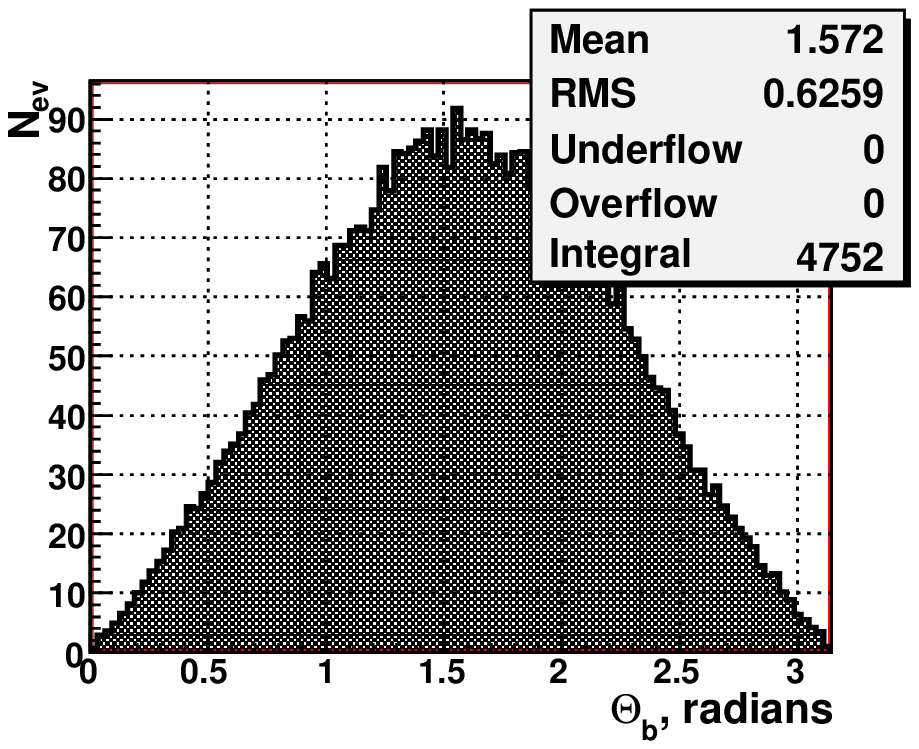}} 
   \mbox{b) \includegraphics[  width=7.2cm, height=5.2cm]{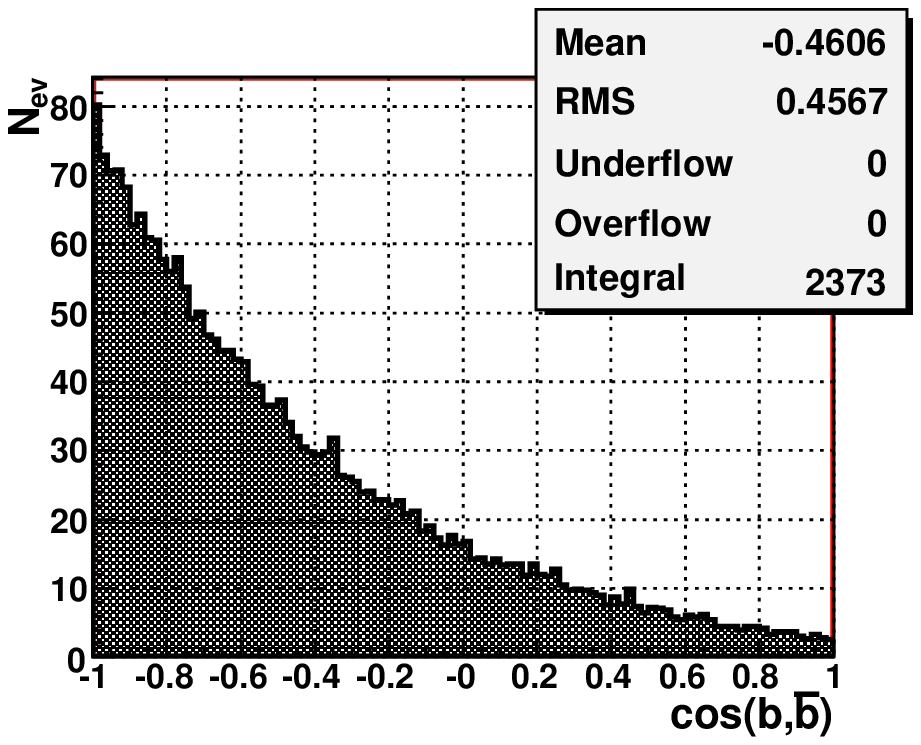}} 
    \end{tabular}
     \caption{\small \it {\bf a)} b- and $\bar b$-quark 
                   polar  angle $\Theta_{b}$ distribution 
		   in stop production; 
		   {\bf b)}$cos(b,\bar b)$  distribution
		   in stop production.} 
     \end{center}   
\vskip -0.5cm            
     \end{figure}

  The distribution of the polar angle  $ \Theta_{b}$ 
   of the $b$-quarks in stop  production
   is presented in plot {\bf a)} of Fig.12.     
   Plot {\bf b)} of
   Fig.12 contains  the  $cos(b,\bar b)$
   distribution, where  $cos(b,\bar b)$  is the cosine 
   of the opening angle between the  3-momenta  of the 
   $b$- and $\bar b$-quarks produced in  the same 
   stop event.  It demonstrates that  most of the
   $b$- and $\bar b$-quarks  move in
   approximately  opposite directions,
   but some  are in the same  hemisphere. 
   Thus in the experiment we may expect a similar 
   angular distributions of the corresponding $b$- and 
   $\bar b$- jets.

       As the next step, we take into account
  $b$-quark  hadronization into a $b$-jet.
  Technically, $b$-jets are defined as jets
  that contain at least  one B-hadron. 
 Their decay may be identified 
  by the presence of a secondary vertex
  \cite{Haw}. 

   Figs.13 and 14 show the energy $E_{b-jet}$  
   and  transverse momentum  $PT_{b-{jet}}$
   distributions  of the corresponding $b$-jets.
   Plots  {\bf a)} and {\bf b)}   are for  stop
   and  top production, respectively.

     \begin{figure}[!ht]
     \begin{center}
    \begin{tabular}{cc}
     \mbox{a) \includegraphics[ width=7.2cm, height=5.2cm]{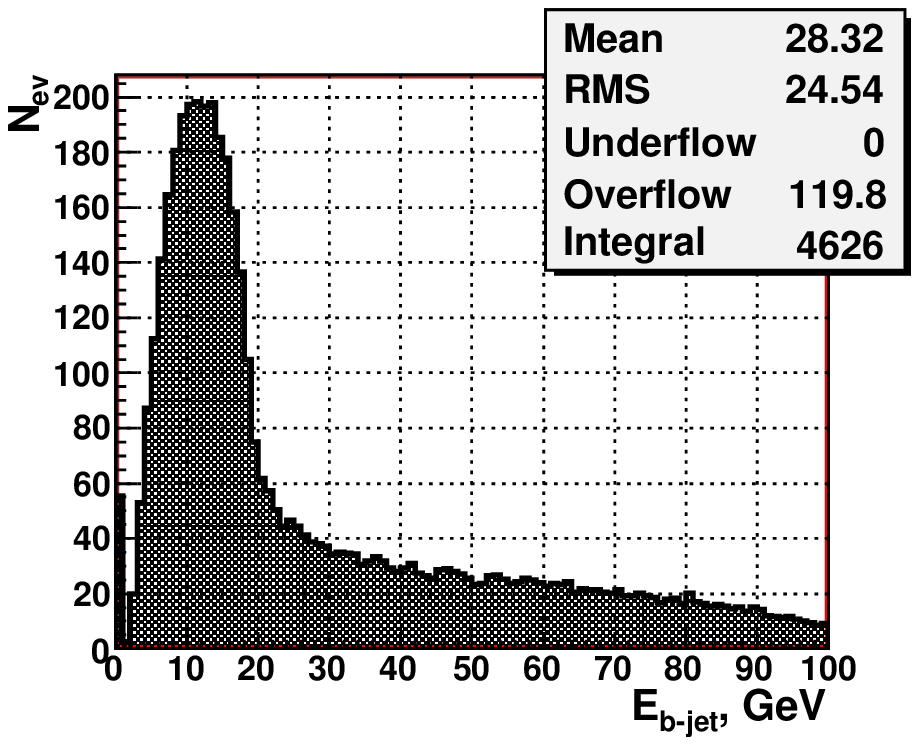}}
     \mbox{b) \includegraphics[width=7.2cm, height=5.2cm]{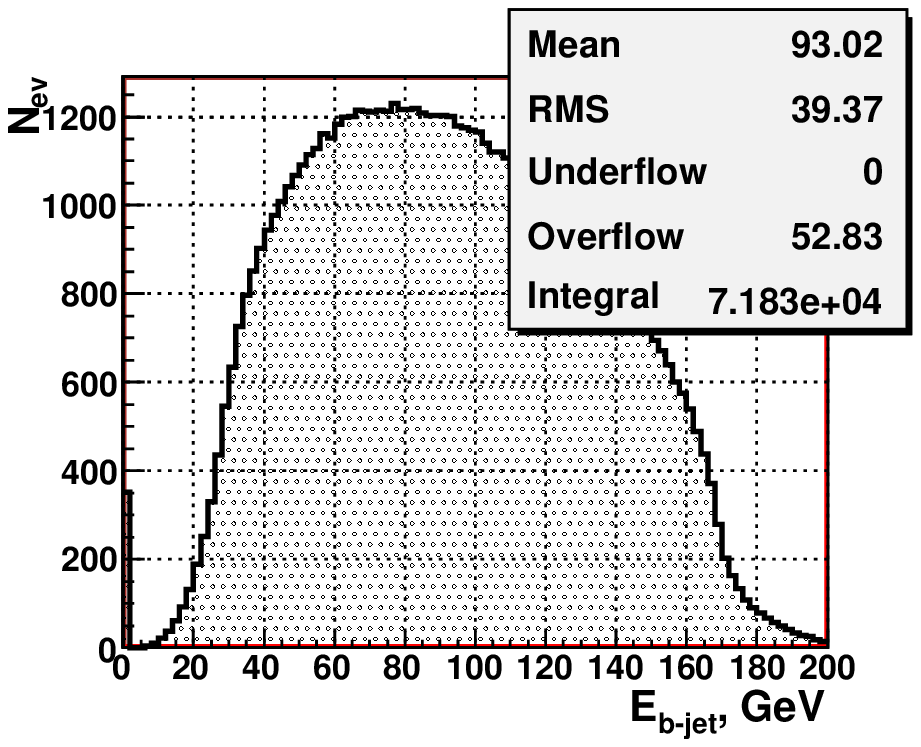}} 
     \end{tabular}
     \caption{\small \it  b-jet energy spectra.
             {\bf a)} stop pair production;
	     {\bf b)} top pair production.}
                         
     \end{center} 
\vskip -0.5cm              
     \end{figure}

      \begin{figure}[!ht]    
    \begin{center}
    \begin{tabular}{cc}
     \mbox{a) \includegraphics[ width=7.2cm, height=5.2cm]{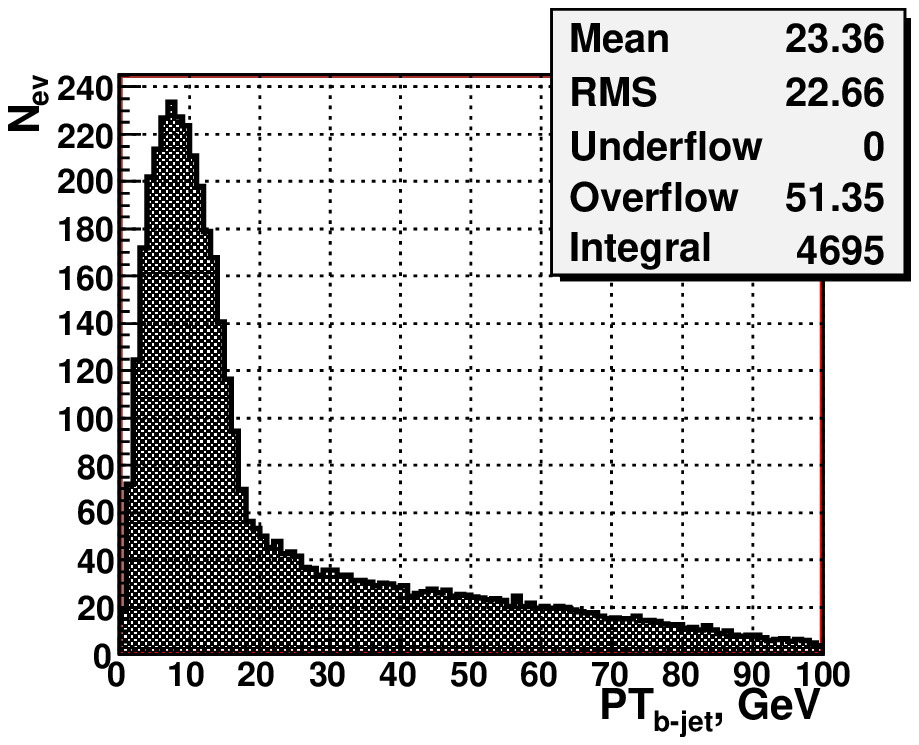}}
     \mbox{b) \includegraphics[  width=7.2cm, height=5.2cm]{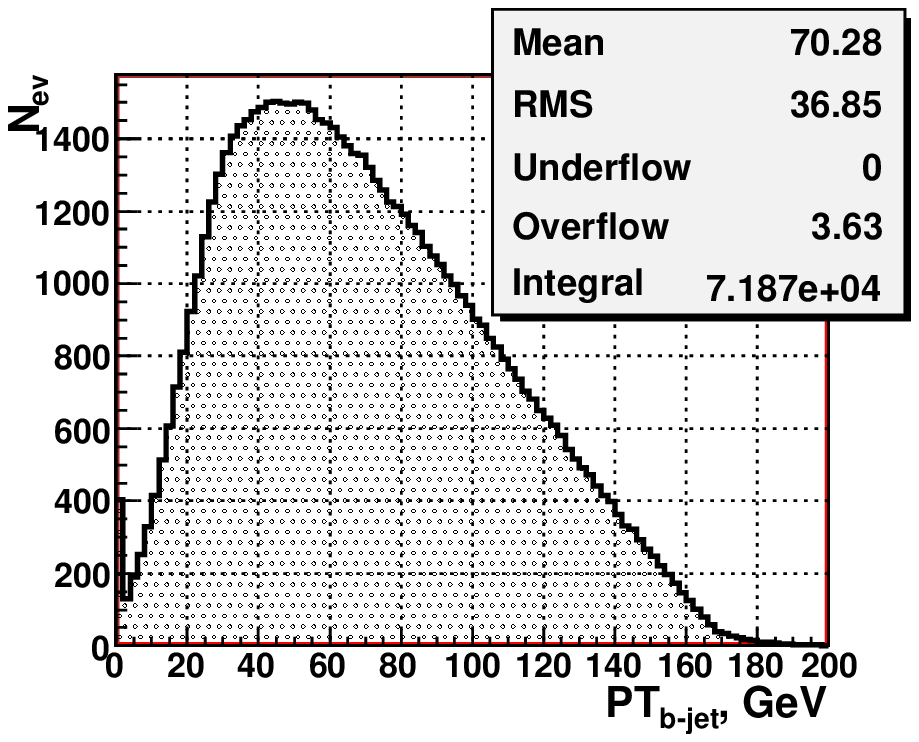}} 
     \end{tabular}
     \caption{\small \it   b-jet  PT- spectra.  
                  {\bf a)} stop pair production;
	          {\bf b)} top  pair production.}                       
     \end{center} 
\vskip -0.5cm              
     \end{figure}
           
   Comparing plots {\bf a)} of Figs.13 and 14
   for the $b$- and  $\bar b$-jet energy
    $E_{b-jet}$  and jet 
   transverse momentum $PT_{b-{jet}}$  
   distributions in  stop production with  
   the  Figs.10 {\bf a)} and 11 {\bf a)},    
   one observes     that the 
   corresponding mean values  of jet
   energy and transverse momentum 
   are about 15-17  GeV higher than those 
   for the quarks. 
  The reason for this is that 
   in the stop case 
   the end points of the energy distributions 
   for the $b$-jets  and  $\bar b$-jets  
   are  higher than those for the
   corresponding  quarks  due to appearance
   of long tails at higher $E_{b-jet}$.

    It is worth emphasizing 
   that at the same time
   the position of the  peak in $b$-jet energy
   distribution $E_{b-jet}$ (see plot {\bf a)}
   of Fig.13) in the stop case
   is only slightly
   shifted to higher values (by about 3-4 GeV)
    compared to the $E_{b}$ peak in plot
   {\bf a)} of Fig 10. Also note  
   that there are  practically no changes of
   the peak positions of the quark level $PT_{b}$ and 
   the jet level $PT_{b-{jet}}$ distributions
   shown in plots {\bf a)} of Figs.11 and
   14, respectively. Thus we can say that the peak
   positions of energy and transverse momentum
   distributions turn out to be rather 
   stable when  passing from $b$-quark 
   to $b$-jet level.     
        
   Now let us turn to the analogous
   energy and transverse momentum 
   distributions obtained for the case of
   top background.
   As seen from comparison of the energy plots
   {\bf b)} of Figs.10 and 13, as well 
   as the transverse momentum
   plots  {\bf b)} of Figs.11 and
    14, the mean values of the 
   $b$-jet and $\bar b$-jet   energy and
   transverse momentum distributions 
   are about 1-3 GeV smaller than the 
   mean values of the corresponding
   $b$-quark and $\bar b$-quark 
   distributions.

   Let us summarize the results which were
   obtained  in subsections  3.2 and 3.3
   by the use of PYCLUS jet finder. First, it was 
   found that in the case of top background
   production the characteristic parameters of
   energy and transverse momentum distributions
   of jets stemming from  W decay and of $b$-jets,
   produced in $b$-quark hadronization, 
   practically do not differ from the
   parameters  of their parent quarks 
   distributions.

   This picture changes quite noticeably
   when we consider the case of stop production 
   with its further decay through the channel
   $\tilde t_{1} \to b \tilde \chi_{1}^{\pm}$. 
   In this case the $b$-quarks
   are much less energetic than the
   b-quarks produced in top decay
   $t \to bW^{\pm}$. It was observed that the
   use of the same PYCLUS jet finder 
   in the stop case leads to a noticeable
   redistribution of jet energies and,
   correspondingly, of their transverse 
   momentum.
   Namely, the  mean values of jet
   energy $E_{jet_W{*}}$ 
   and jet transverse momentum 
   $PT_{jet_{W^{*}}}$ are about 15-17 GeV 
   smaller  than the  energy $E_{W-quark}$ and  
   transverse momentum
   $PT_{W-quark}$  of parent  "$W$-quarks"
   (stemming from W boson decay), while the mean
   values of $b$-jet energy $E_{b-jet}$ and jet 
   transverse momentum $PT_{b-{jet}}$ are
   by about 15-17 GeV higher than the energy
   $E_{b}$ and $PT_{b}$  of parent $b$-quarks.
   
   In the following we shall return to this
   subject and consider the set of physical
   variables which shall take into account
   this effect of energy redistribution  in a 
   case of stop production.

%
    \section{ Distributions of the signal muons.}
%
  ~~~ To select the signal stop pair
   production events  shown  in the left
   diagram of Fig.1 one has to identify the
   muon from the W decay. The corresponding
   energy $E_{sig-mu}$ and transverse
   momentum $PT_{sig-mu}$ distributions 
   of the signal muons are shown  in  Fig.15.
     
  \begin{figure}[!ht]
     \begin{center}
    \begin{tabular}{cc}
    \mbox{a) \includegraphics[  width=7.2cm, height=5.2cm]{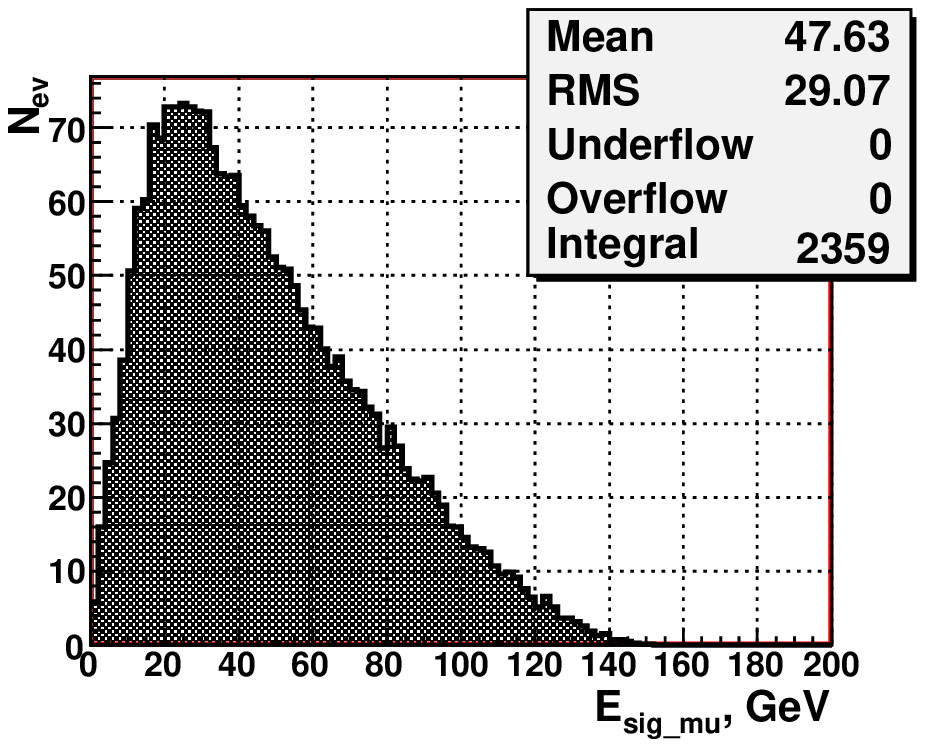}}      
    \mbox{b) \includegraphics[  width=7.2cm, height=5.2cm]{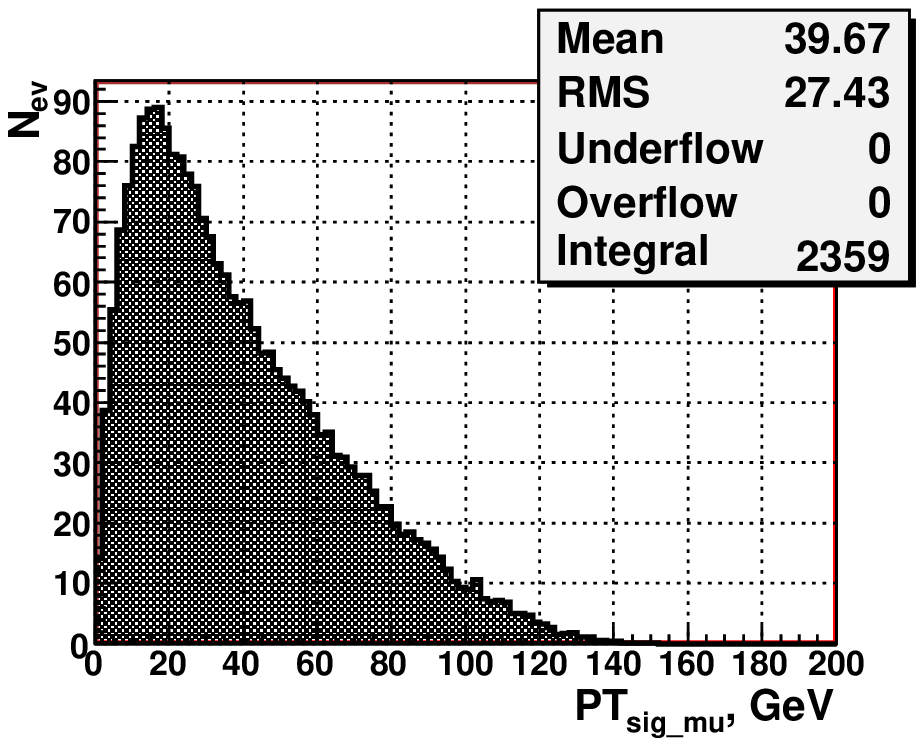}}      
     \end{tabular}
     \caption{\small \it {\bf a)} Energy distributions 
                        of signal muons. 
                        {\bf b)} PT distributions 
                        of signal muons.} 
     \end{center} 
  \vskip -0.5 cm             
     \end{figure}

      There are, however, also muons in the 
   event coming from
   leptonic and semileptonic decays of hadrons. 
    Fig.16 {\bf a)} and  {\bf b)} 
   show, respectively, the energy 
   $E_{dec-mu}$ and the  transverse momentum
   $PT_{dec-mu}$ spectra of these
   background  muons stemming from hadron 
   decays within the detector 
   volume (for which we took the size 
   from \cite{ILCRDR1}, \cite{ILCRDR2}). 
    It can be  seen that the decay muons 
   have a rather small energy 
   $E_{dec-mu}$ and transverse momentum 
   $PT_{dec-mu}$. Their mean values are 
   about $0.79$ and $0.63$ GeV,
   respectively. The analogous spectra 
   for the signal muons  in Fig.15  
   show that the  signal muons  have a 
   much higher energy   $E_{sig-mu}$ and 
   transverse momentum  $PT_{sig-mu}$. 
   The mean value of the signal muons  
   energy 
    $E^{mean}_{sig-mu} = 47.6$ GeV 
   is about 60 times higher than the 
   mean value of the energy
   of the decay muons. An analogous
   difference can be seen between the 
   mean values of transverse momenta
   PT of signal and decay muons. 
      One can cut off
   most low--energy decay muons rejecting 
   those  with  $E_{mu} \le 4$ GeV. Such
   a cut leads to a loss of about 15-20 
   signal events as seen from the 
   plot $\bf a)$ of Fig.15
   (the bin width in this plot is 2 GeV).
 
 \begin{figure}[!ht]
     \begin{center}
    \begin{tabular}{ccc}
     \mbox{a) \includegraphics[   width=7.2cm, height=5.2cm]{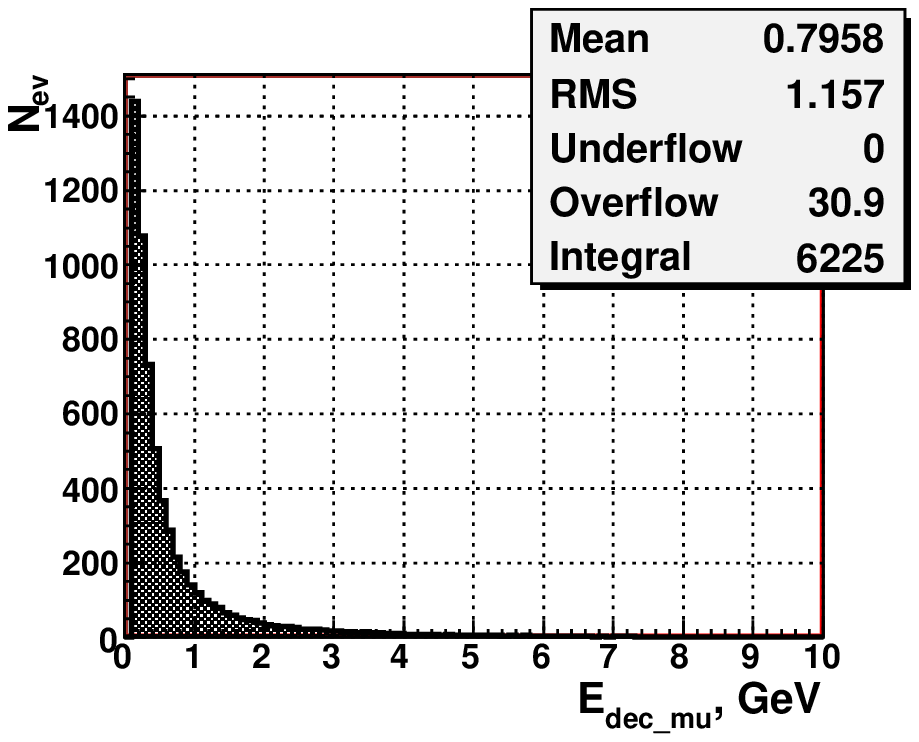}}      
     \mbox{b) \includegraphics[  width=7.2cm, height=5.2cm]{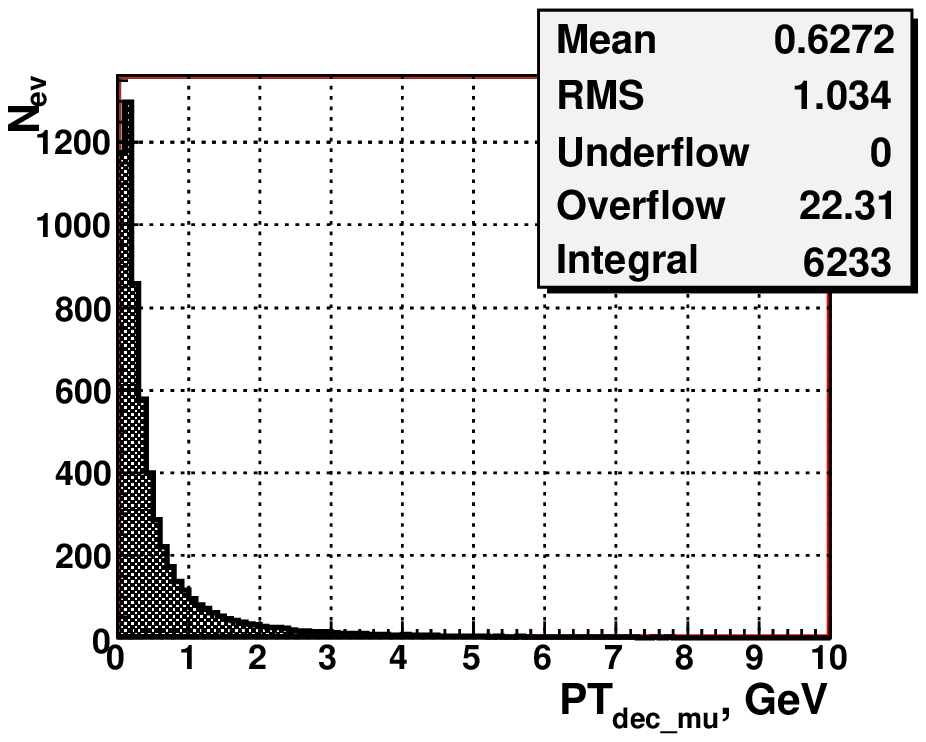}} 
    \end{tabular}
     \caption{\small \it 
     Distributions of background muons. 
     {\bf a)} Energy distribution; {\bf b)}
              PT distribution.}    
     \end{center} 
\vskip -0.5cm             
     \end{figure}

   We have also studied another way to 
   select the signal muon from W decay. 
   If the axes of all four jets in the 
   event are known, then in general 
   the signal muon has the largest
   transverse momentum with respect 
   to any of  these jet axes.

%
\section{Global variables  useful for
        background separation.}
%

~~~~~In  stop pair production the  two 
  neutralinos and  the energetic neutrino 
  from the W boson decay escape detection. 
  The simulation with PYTHIA6 allows us 
  to estimate the missing energy and the
  missing transverse momenta that are
  carried away by these particles.
   We also take into account the
  non-instrumented region around the beam 
  pipe given by the polar angle intervals 
   $\Theta < 7^o$ and  $\Theta > 173^o$.

  The distributions of the  total missing 
  $E_{miss-tot}$ energy for stop production
  and  top production are presented in
  {\bf a)} and  {\bf b)} plots of Fig.17,
  respectively. In  stop pair production,
  see plot {\bf a)}, the  $E_{miss-tot}$ 
  spectrum starts at 200 GeV.
  In  top pair production (plot {\bf b)}),
  where two neutralinos are not present, 
  the missing energy $E_{miss-tot}$ is 
  much smaller, going from 
  $ \approx 10$ GeV to $ \approx 260$ GeV.
  
     \begin{figure}[!ht]
     \begin{center}
    \begin{tabular}{cc}    
  \mbox{a) \includegraphics[   width=7.2cm, height=5.2cm]{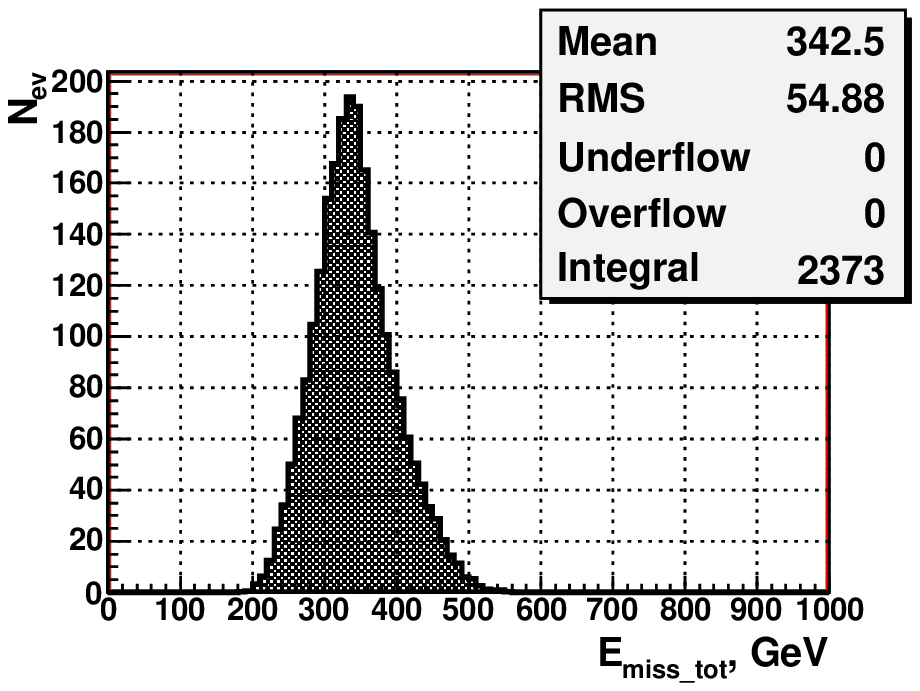}}
     \mbox{b) \includegraphics[   width=7.2cm, height=5.2cm]{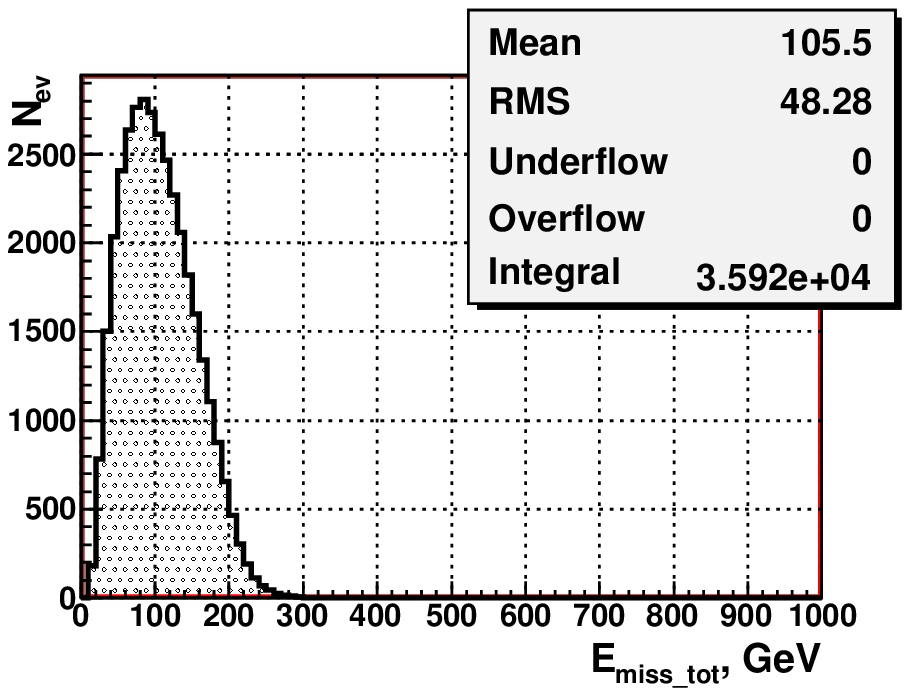}}      
    \end{tabular}
     \caption{\small \it Missing energy $E_{miss-tot}$  
                         distribution.  
	              {\bf a)} stop pair production;
	              {\bf b)} top pair production.} 
\end{center} 
\vskip -0.5 cm           
    \end{figure}
     
    Figure 18 shows the distributions of 
  the total visible energy 
  in the event, $E_{vis-tot}$, 
  in stop production (plot {\bf a)}) and in
  top  production (plot {\bf b)}). The large 
  missing energy in  stop production (Fig.17)
  is related to the low visible energy 
  (Fig.18), while in  top production 
  the low missing energy correlates with 
  the large visible energy. A cut on the 
  total visible energy of approximately 
  $E_{vis-tot} < 220 $ GeV
  \footnote{That is equivalent to setting a 
          lower limit for the missing energy.}
  would  eliminate almost completely the top 
  background, while leaving the most
  part of the signal events.
  
       \begin{figure}[!ht]
     \begin{center}
    \begin{tabular}{cc}
     \mbox{a) \includegraphics[width=7.2cm, height=5.2cm]{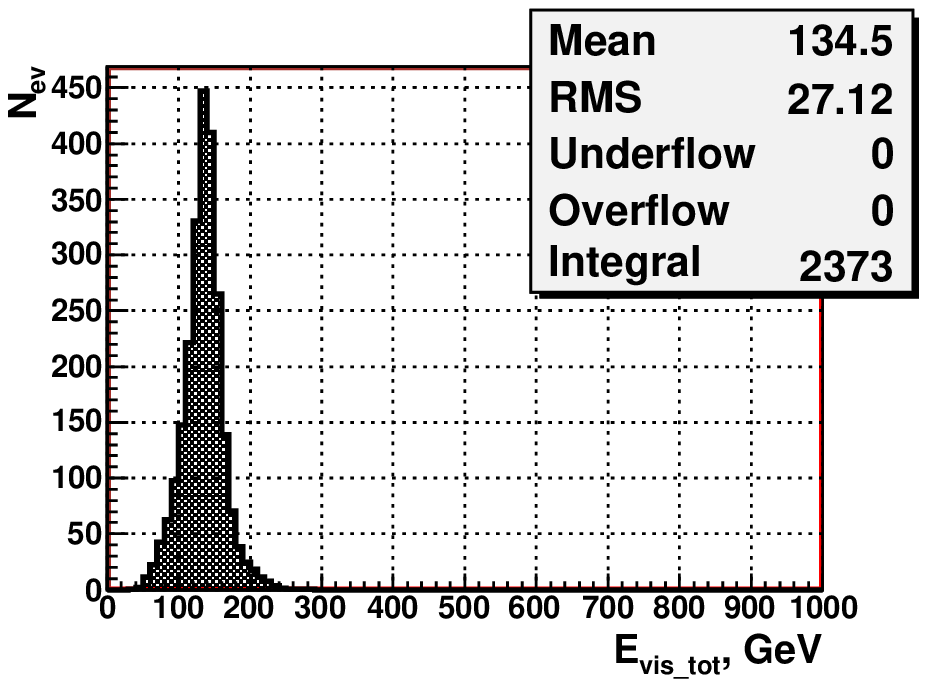}} 
     \mbox{b) \includegraphics[  width=7.2cm, height=5.2cm]{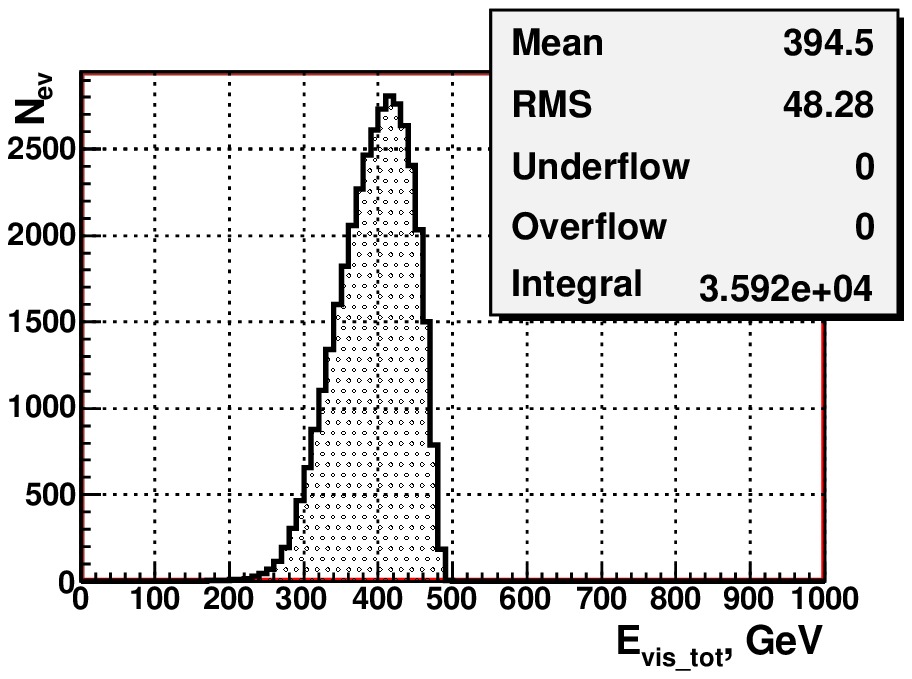}}      
    \end{tabular}
     \caption{\small \it Total visible energy 
             $E_{vis-tot}$ distribution. 
	         {\bf a)} stop  pair production;
	         {\bf b)}  top  pair production.}            
     \end{center}  
\vskip -.5 cm           
     \end{figure}

   Another useful observable is the scalar
   sum  of the  moduli of the transverse
   momenta in an event  $PT_{scalsum} =
   \sum\nolimits_{i=1}^{N^{part}} |PT_{i}|$,
   where the sum goes over all ($N^{part}$)
   detectable particles in the  event. 
   Figure 19 shows the distributions of 
   the scalar sum of the transverse momenta 
   for the stop production (plot {\bf a)})
   and for  top production (plot {\bf b)}). 
   It is seen that the restriction
   $PT_{scalsum} \leq 150$ GeV  would lead
   to a good separation of the stop signal
   events from the top background.

   \begin{figure}[!ht]
     \begin{center}
    \begin{tabular}{cc}
    \mbox{a) \includegraphics[  width=7.2cm, height=5.2cm]{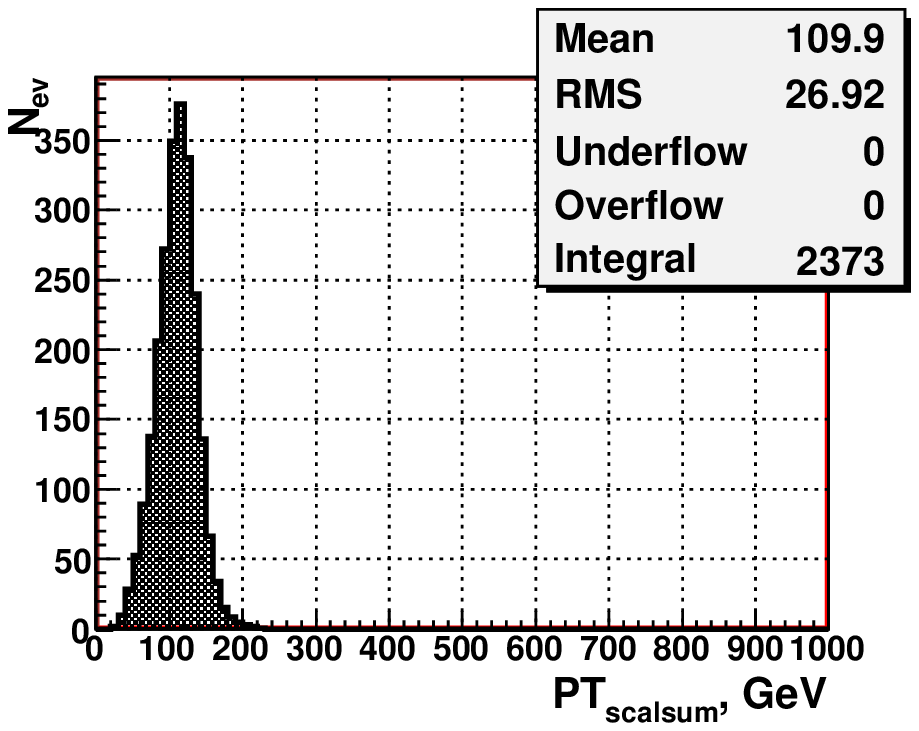}} 
     \mbox{b) \includegraphics[ width=7.2cm, height=5.2cm]{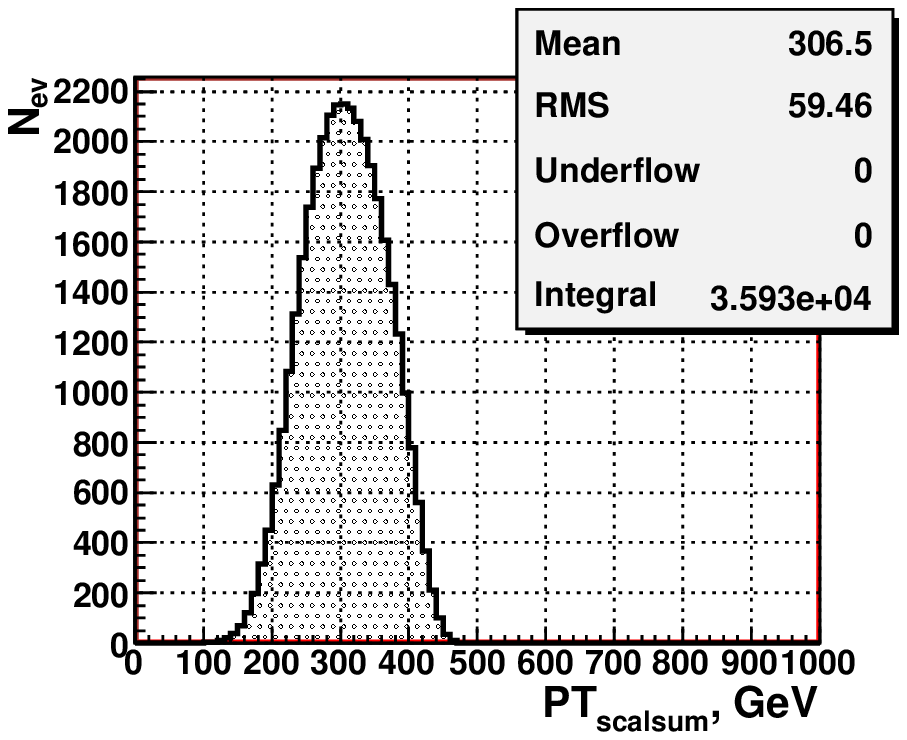}}      
    \end{tabular}
     \caption{\small \it $PT_{scalsum}$- distribution.
                   {\bf a)}  stop pair  production;
	           {\bf b)} top pair production.} 
             
     \end{center}
 \vskip -0.5 cm            
     \end{figure}

  Let us 
  consider also the invariant mass of the 
  system that contains all observable 
  objects in the final state. This invariant
  mass is the modulus of the  vectorial 
  sum  of the  4-momenta  $P^{n}_{jet}$ 
   \footnote{ The 4-momentum 
        $ P^{n} =(P^{n}_{0},  \bf P^{n})$,
        of the n-th jet includes its energy 
	$P_{0}^{n} = E^{n}$ and the components
	of the 3-dimensional momentum 
   ${\bf P}^{n}$
   $=(P^{n}_{x},P^{n}_{y},P^{n}_{z})$.}
   of  all  ($N^{jet} = 4, n=1,2,3,4 $) 
   jets in an event 
   plus the 4-momentum of the signal 
   muon $P_{\mu}$,

\begin{equation}  
  M_{inv}(All jets, mu) = 
  \sqrt{(\sum\nolimits_{n=1}^{N^{jet}}{P^{n}_{jet}+P_{\mu})^{2}}} .
\end{equation} 
  
  The distribution of this invariant mass 
  is shown in Fig.20.  Plot {\bf a)} 
  shows the results  for   stop pair 
  production  while plot {\bf b)} is 
  for  top pair production. As seen from
  these plots,  the cut 
   $M_{inv}(All jets, \mu) \le 200$  GeV 
  will give a  good separation of signal 
  stop and top background events.
 
 \begin{figure}[!ht]
     \begin{center}
    \begin{tabular}{cc}
 
     \mbox{a) \includegraphics[ width=7.2cm, height=5.2cm]{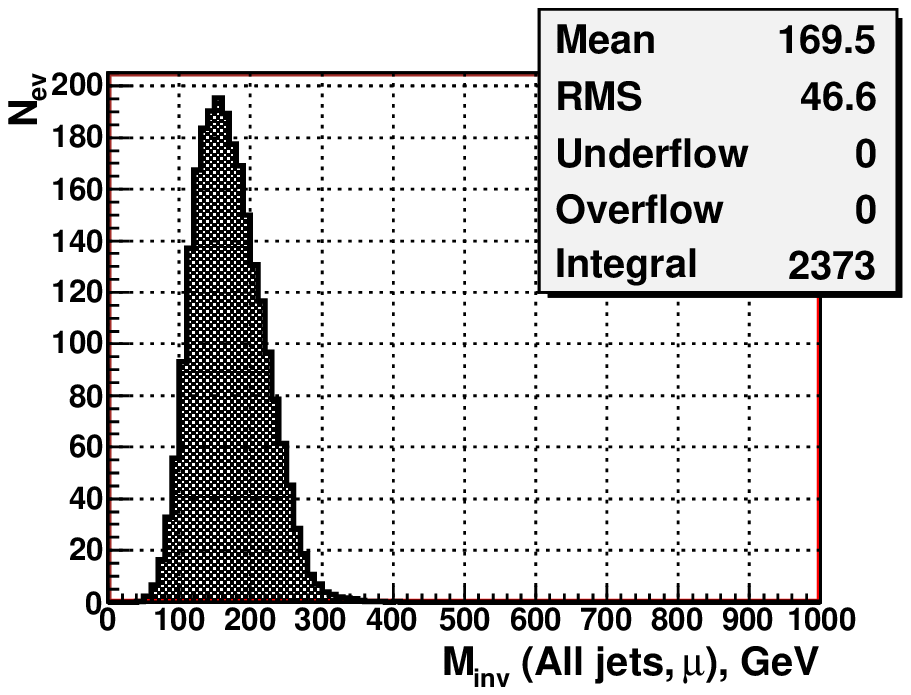}}      
     \mbox{b) \includegraphics[ width=7.2cm, 
height=5.2cm]{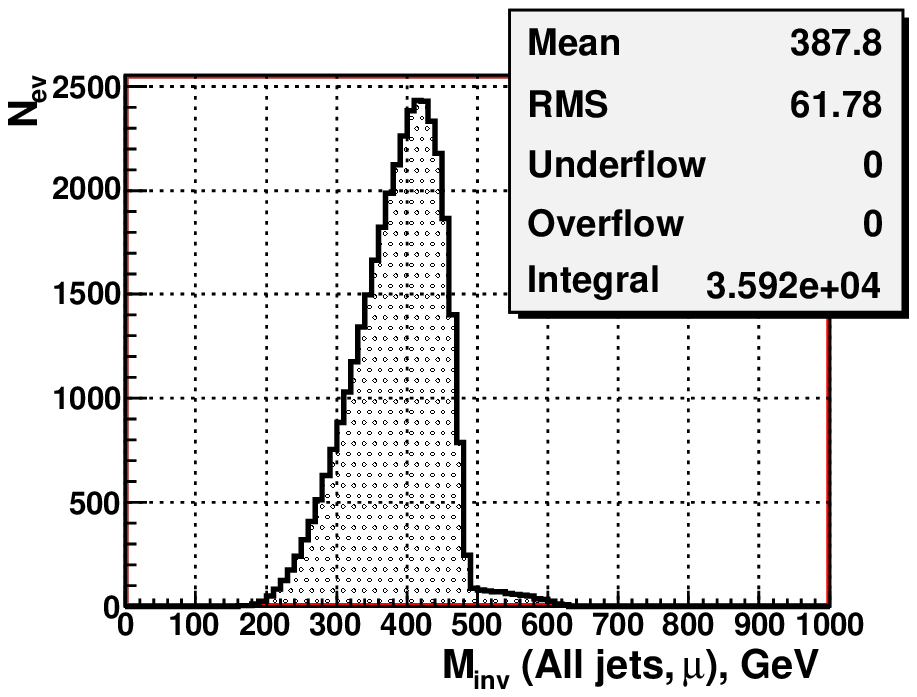}}      
     \end{tabular}
     \caption{\small \it Distribution of number
                 of events, 
		 $N_{event}$ $(L_{int}=1000 fb^{-1})$,
		 versus the  reconstructed 
		 invariant mass of all jets and 
		 signal muon $M_{inv}(All jets, mu)$.
	       	 {\bf a)} stop pair production;
	         {\bf b)} top pair production.}
     \end{center}    
\vskip -0.5cm           
     \end{figure}

%
%
    \section{ Global jet variables and cuts.}
%
%

   ~~~ An even more efficient separation of
   the signal and  the background can be
   obtained by using  the   invariant mass
   of all jets $M_{inv}(All jets)$ 
   which is the modulus of the
   vectorial sum  of the  4-momenta of  all
   $N^{jet}$ jets in an event 
\begin{equation}    
   M_{inv}(All jets)=
   \sqrt{(\sum\nolimits_{n=1}^{N^{jet}}{P^{n}_{jet})^{2}}} .
\end{equation}    

   The distribution of this invariant mass
  is shown in Fig.21.   Plot {\bf a)} 
  presents  the result for stop
  pair production  while plot {\bf b)} that
  for top pair production.  It is seen
  that the application of  the cut 
  $M_{inv}(All jets) \le 160$ GeV leads to 
  practically a complete separation of signal
  stop and  top background  events.

 \begin{figure}[!ht]
     \begin{center}
    \begin{tabular}{cc}
     \mbox{a) \includegraphics[   width=7.2cm, height=5.2cm]{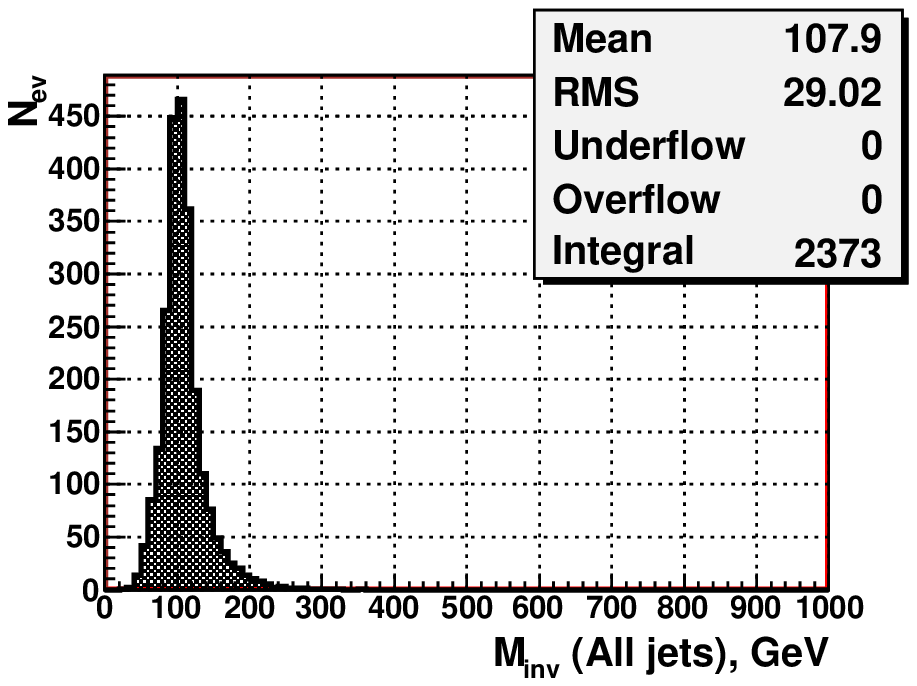}}      
     \mbox{b) \includegraphics[    width=7.2cm, height=5.2cm]{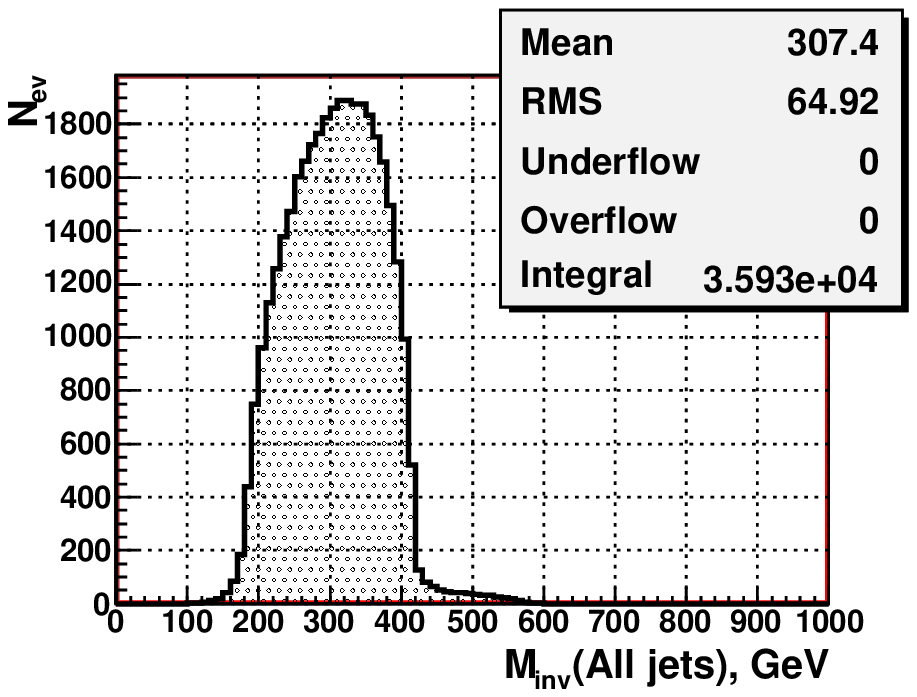}}      
     \end{tabular}
     \caption{\small \it Distribution of number 
                 of events,
		 $N_{ev}$ $(L_{int}=1000 fb^{-1})$,
		  versus the reconstructed 
		 invariant mass of all jets
                 $M_{inv}(All jets)$.
		   {\bf a)} stop pair production;
	           {\bf b)} top pair  production.}
     \end{center}  
\vskip -0.5cm              
     \end{figure}

   Another  variable that can also be used 
   for the separation of the signal  and  
   the background is the "missing" mass 
   $ M_{missing}$ 
   (we use  $\sqrt {s}=500$ GeV)
 \begin{equation}  
  M_{missing} = 
   \sqrt{ (\sqrt {s} -
   (\sum\nolimits_{n=1}^{N^{jet}}{E^{n}_{jet}+
            E_{\mu}))^{2}} -
   (\sum\nolimits_{n=1}^{N^{jet}}{{\bf P}^{n}_{jet}+
            \bf P_{\mu})^{2}} }
\end{equation} 

  This variable takes  into account  the 
  contribution of  those particles that 
  cannot be registered in the detector 
  (neutrinos and neutralinos). 
  The distributions of this invariant
  "missing" mass are given in Fig.22. 
  Plot {\bf a)} shows the results  for   
  stop pair  production  while plot {\bf b)}
  is for top pair production.  As can be seen 
  from these plots, the cut 
   $M_{missing} \ge 250$ GeV
 also allows us to get  rid of  most
  of the top background contribution.

\begin{figure}[!ht]
     \begin{center}
    \begin{tabular}{cc}
     \mbox{a) \includegraphics[  width=7.2cm, height=5.2cm]{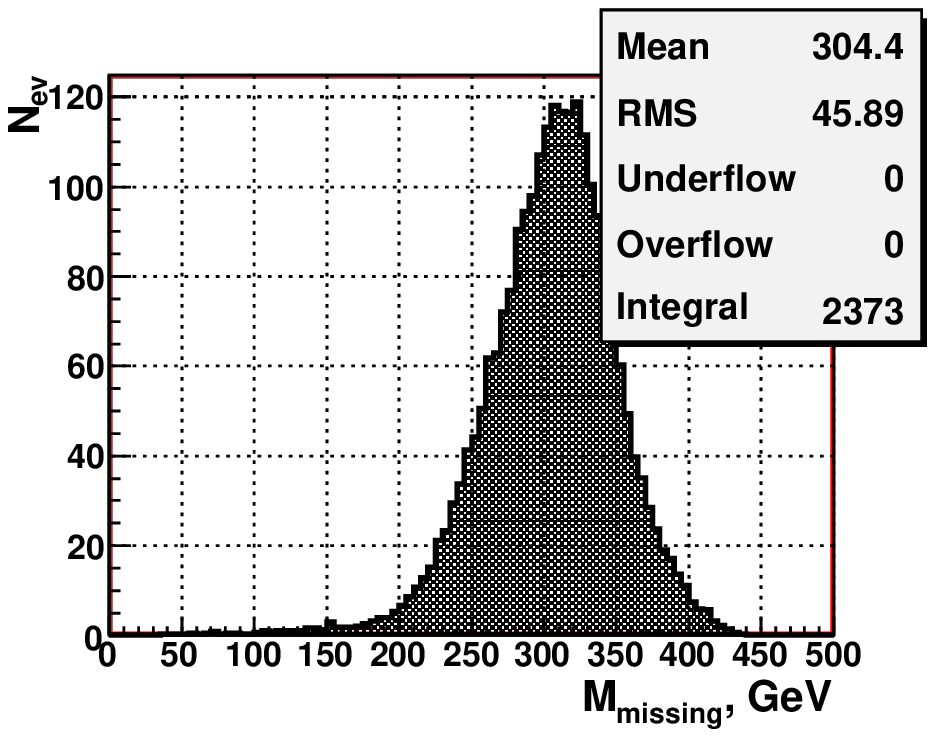}}      
     \mbox{b) \includegraphics[ width=7.2cm, height=5.2cm]{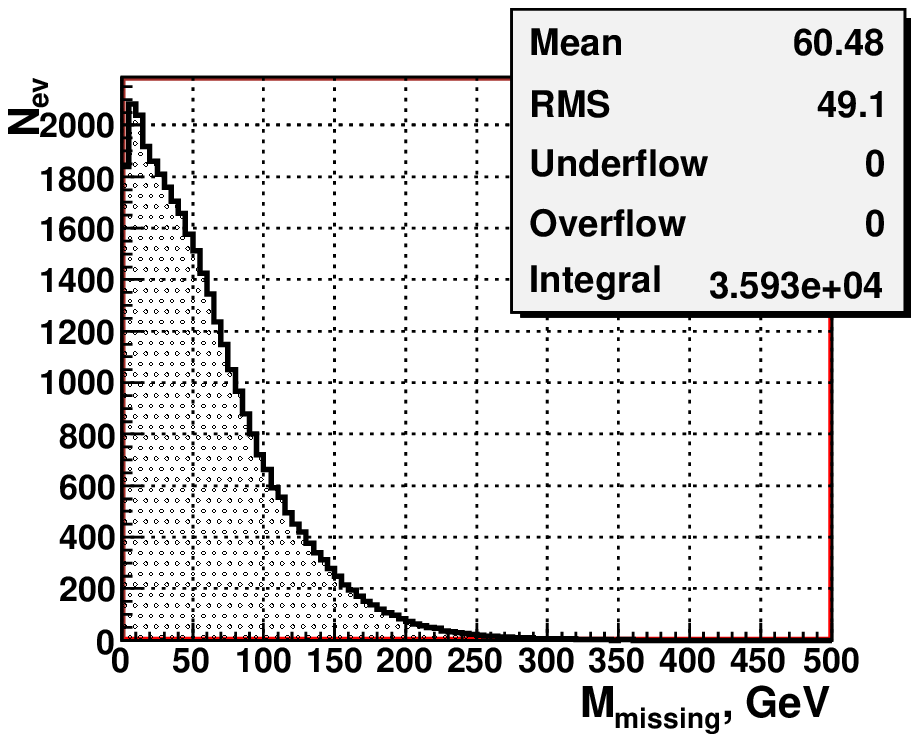}}      
     \end{tabular}
     \caption{\small \it Distribution of number 
                 of events,
	    $N_{ev}$ $(L_{int}=1000 fb^{-1})$, 
             versus the missing mass variable.
            {\bf a)} stop pair production;
	    {\bf b)} top pair  production.}
     \end{center}   
\vskip -0.5cm                  
     \end{figure}

%
\subsection{ Cuts and signal-to-background ratio.} 
%

~~ To diminish the influence of jet energy
   redistribution effect, discussed in subsections
   3.2 and 3.3,
   we shall use the cuts considered above for 
  $ M_{missing}$ and 
   $M_{inv}(All jets)$. These variables, by
   definition, include the total 4-momentum of
   all jets, defined as the  vectorial 
   sum  of the  4-momenta of all jets. Therefore
   they do not suffer from energy redistribution
   between jets. 
   Consequently, we use the
   following three cuts to separate the 
   signal and background events:\\

$\bullet$ there must be at least two $b$-jets in an event: 

\begin{equation} 
     N_{b-jets} \geq 2 ; 
\end{equation}
$\bullet$ the invariant missing mass must
               be larger than $250$ GeV:
\begin{equation}   
              M_{missing}  \ge  250 ~GeV;
\end{equation}
$\bullet$ the invariant mass of  all jets  must
             be smaller than  $160$ GeV:
\begin{equation}   
               M_{inv}(All jets) \leq 160 ~GeV.
\end{equation}
	   
These three cuts reduce the number of top background events from $\approx 3.5*10^4$ 
  to    $\approx 12$  and the number of stop signal
  events from 2373 to 1806.    
  So, the cuts improve the 
  signal--to--background  ratio from $S/B=0.066$
  to $S/B \approx 143$ 
  losing about 
   $23.8\%$  of the signal stop events.
  The efficiency values for the  cuts (14)--(16)
  are calculated for $\sqrt {s}=500$ GeV. 
  We define them  as
  the summary efficiencies. It means that if 
   $\varepsilon_{1}$ 
  is the  efficiency of the first cut (14), 
   $\varepsilon_{12}$   is  the 
  efficiency of applying the first cut (14) 
  and then second cut (15).
   Analogously,  $\varepsilon_{123}$
   is the efficiency of  the successive 
   application of the cuts (14), (15) and (16).
 The following results are obtained:  \\
   
  For SIGNAL STOP events: 
~~~~~~~~~ $\varepsilon_{1}  =0.84$; ~~ $\varepsilon_{12}  =0.78$; ~~~~$ \varepsilon_{123}= 0.76$;

 For BACKGROUND TOP events : 
 $\varepsilon_{1} =0.94$; ~~ $\varepsilon_{12}=0.001$; ~~ $\varepsilon_{123}=  3.3  \cdot 10^{-4}$. \\

%
\subsection{ Signal and background cross sections.}
%
 ~~~~ We give  the cross
 sections and the  numbers of events for  stop pair
  production  and top pair production for five 
  energies  $\sqrt {s}$ in Table 1 without cuts and Table 2 
  with cuts. To calculate the rates of 
  stop production
  at different  energies  we take 
  the   value of the integrated
  luminosity $L_{int}=$ 1000 $fb^{-1}$
  for all energies.
  
  It is seen that the  cuts (14)-(16)
 lead to a strong suppression (more than 3 orders of
 magnitude)   of the background top 
 contribution and a moderate loss of  
 signal stop events produced in the energy
 range  $400 \leq \sqrt {s} \leq 800$ GeV.  

  For the stop mass chosen, the largest
  number of signal events 
  is  expected   at  $\sqrt {s}=500$ GeV.   
  Let us note that, according to  
   \cite{Damerell}, a  $50\%$ efficiency of
   the separation of $b$ -jets
   and  $80\%$ of the  corresponding 
   purity can be  expected.
   It means that  in reality  to get
   1806  signal stop events,  reconstructed
   with an account of $b$- tagging,  we
   will  need about 2.5 times higher 
   statistics than that  provided
   by  the  integrated luminosity  1000 fb$^{-1}$
   at  $\sqrt {s}=500$ GeV. 
 
   It is worth noting  that,  as seen from
  Table 2,  even with the use 
  of the fixed parameters of the  cuts 
  (14)--(16) the number of signal stop  
  events that can  pass these cuts  grows
  rapidly  with  the energy in the region
  $400 \leq \sqrt {s} \leq 500$ GeV where 
  mass measurements will
  be done in the first phase.
  These  measurements may allow to enlarge the
  sample of collected signal events and to
  perform   a precise measurement of the stop mass.
   The region  $500 \leq \sqrt {s} \leq 800$ GeV,
   which will   be available in the second phase of
   ILC operation,  allows to gain  a much higher stop statistics, as
  seen in   Table 2.
  A complete  analysis based on adjusting  the 
  parameters of the selection cuts (15)-(16) 
  for each of the above energies intervals
   will be presented in   forthcoming papers.

\begin{table}[ht] 
 \vskip -0.5 cm 
 \caption{ The cross sections and the number of events
           for stop and top pair production before
           cuts.
          $L_{int}=1000 fb^{-1}$ is assumed
          for each energy}. 
\begin{center} 
 \begin{tabular}{||c||c|c||c|c||c||}\hline\hline
  
  $2E_{b}=\sqrt {s} ~~[GeV]$ & 
 $\sigma^{e^{+}e^{-}}_{stop} ~~[fb]$ &
 $ N_{stop}^{events} $ & 
 $\sigma^{e^{+}e^{-}}_{top} ~~[fb]$ & 
 $ N_{top}^{events} $  &
 $ S / B $\\ \hline \hline

 350 & 0.23 & 233 & 13.76 & 13750 & 0.0169 \\
 400 & 1.34 & 1347 & 38.79 &  38740 & 0.0347 \\
 500 & 2.37 & 2378 & 35.94 &  35950 & 0.0661 \\
 800 & 1.89 & 1809 & 17.36 & 17359 & 0.1042 \\
 1000 &  1.42 & 1265 & 11.66 & 11656 & 0.1085 \\
 \hline \end{tabular}
\end{center}
 \vskip -0.5 cm  
 \end{table}
 
\begin{table}[ht] 
 \vskip -0.5 cm  
\caption{ The same as in Table 1  but after
          applying cuts (14)-(16).}
\begin{center}   
 \begin{tabular}{||c||c|c||c|c||c||}\hline\hline
 
  $2E_{b}=\sqrt {s} ~~[GeV]$ & 
 $\sigma^{e^{+}e^{-}}_{stop} ~~[fb]$ &
 $ N_{stop}^{events} $ & 
 $\sigma^{e^{+}e^{-}}_{top} ~~[fb]$ & 
 $ N_{top}^{events} $  &
 $ S / B $\\ \hline \hline

 350  &  0.0089 & 8    & 0                & 0 &   \\
 400  &  0.52   & 521  & 2.32 * $10^{-4}$ & 0.2 &  2605    \\
 500  &  1.80   & 1806 & 2.26 * $10^{-2}$ & 12.6  &  143    \\
 800  &  0.99   & 995  & 1.08 * $10^{-2}$ & 10  &  99      \\
 1000 &  0.41   & 410  & 6.26 * $10^{-3}$ & 6   &  69      \\
 
 \hline \end{tabular}
\end{center}
\vskip -0.5 cm  
  \end{table}

%
%
    \section{Determination of scalar top quark mass.}  
%
%

 ~~~ Another variable of interest is  the 
  invariant   mass 
  $M_{inv}$($b-jet$, $JETS_{W}$):
     \footnote{ We follow here the notations 
                of subsections 3.2 and 3.3} 
\begin{eqnarray}
   M_{inv}(b-jet,JETS_{W}) \equiv  
    \sqrt{(P_{b-jet} +P_{JETS_{W}})^{2}},
\end{eqnarray}
  which is constructed as the modulus of the
  vectorial sum of the 4-momentum   $P_{b-jet}$    
    of the $b$-jet, plus the total
    4-momentum of $JETS_{W}$ system, 
     i.e., non-$b$-jets  stemming from the W decay
($P_{JETS_{W}}= P_{jet1_{W}}
       + P_{jet2_{W}}$, as there are only two   jets allowed 
 to be produced in W decay). 
    More precisely, if the signal event 
    contains a $\mu^{-}$ as the signal  muon
     (see Fig.1), we 
    have to take the  $b$-jet ($\bar{b}$-jet
    in the case of $\mu^{+}$ 
    as the signal muon). This 
    is only possible if one
    can discriminate between the $b$- and 
    $\bar{b}$-jets experimentally.
    Methods of experimental determination
    of the  charge of the
    $b$-jet ($\bar{b}$-jet) were developed in  
     \cite{Damerell}. In the present paper we 
    do not use any b-tagging 
    procedure. The PYTHIA information  about
    quark flavor  is taken for
    choosing the $b$- and $\bar{b}$-jets.
 
     The distributions of the  invariant mass of the
  "$b$-jet+$JETS_{W}$"  system is shown in  Fig.23.
  Their analogs $M_{inv}(b, 2~quarks_W)$,
  obtained at quark level, are presented in Fig.24.
  Plots {\bf a)} of these two Figures  show the
  results  for   stop pair 
  production  while plots {\bf b)} are for 
  top pair production.  These distributions
are obtained   without  any cuts.

   \begin{figure}[!ht]
     \begin{center}
    \begin{tabular}{cc}
     \mbox{a) \includegraphics[ width=7.2cm, 
height=5.2cm]{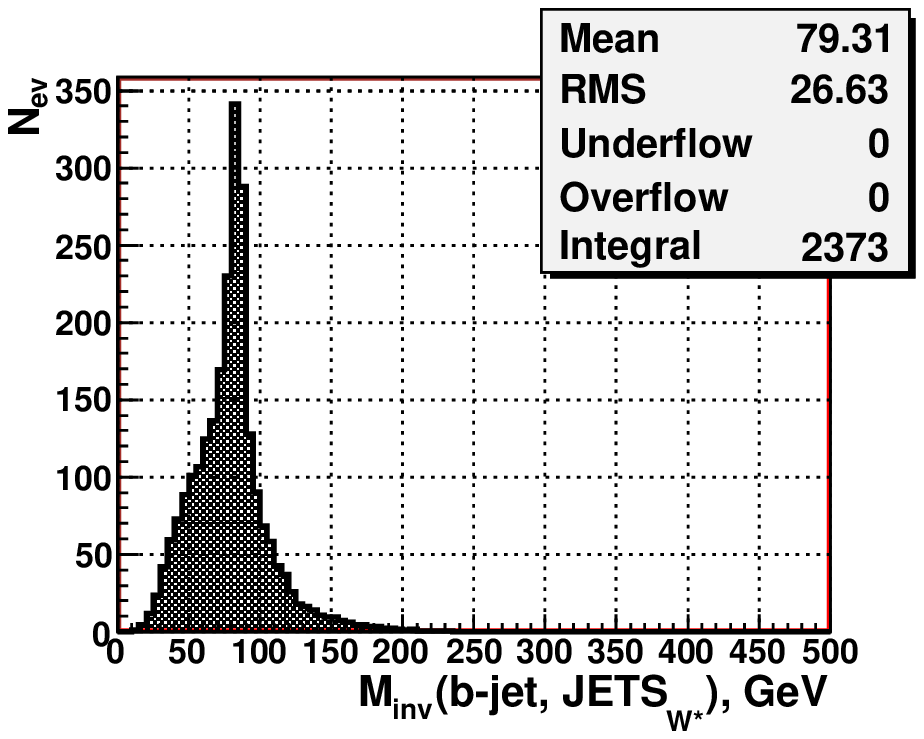}}      
     \mbox{b) \includegraphics[  width=7.2cm, height=5.2cm]{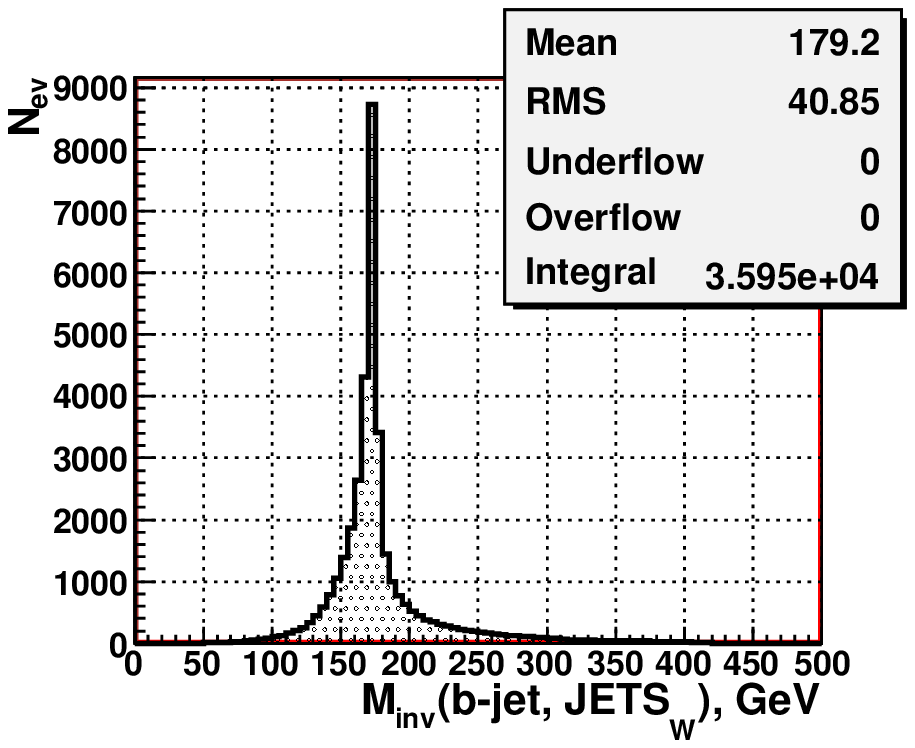}}      
    \end{tabular}
     \caption{\small \it The spectra of the
                        invariant masses
	                $M_{inv}$($b-jet$, $JETS_{W}$)
                        before the cuts (14) - (16).
                        {\bf a)}  stop pair  production;
	                {\bf b)} top  pair production.}
     \end{center} 
\vskip -0.5cm             
    \end{figure}        
     
   \begin{figure}[!ht]
  \begin{center}
   \begin{tabular}{cc}
     \mbox{a) \includegraphics[width=7.2cm, height=5.2cm]{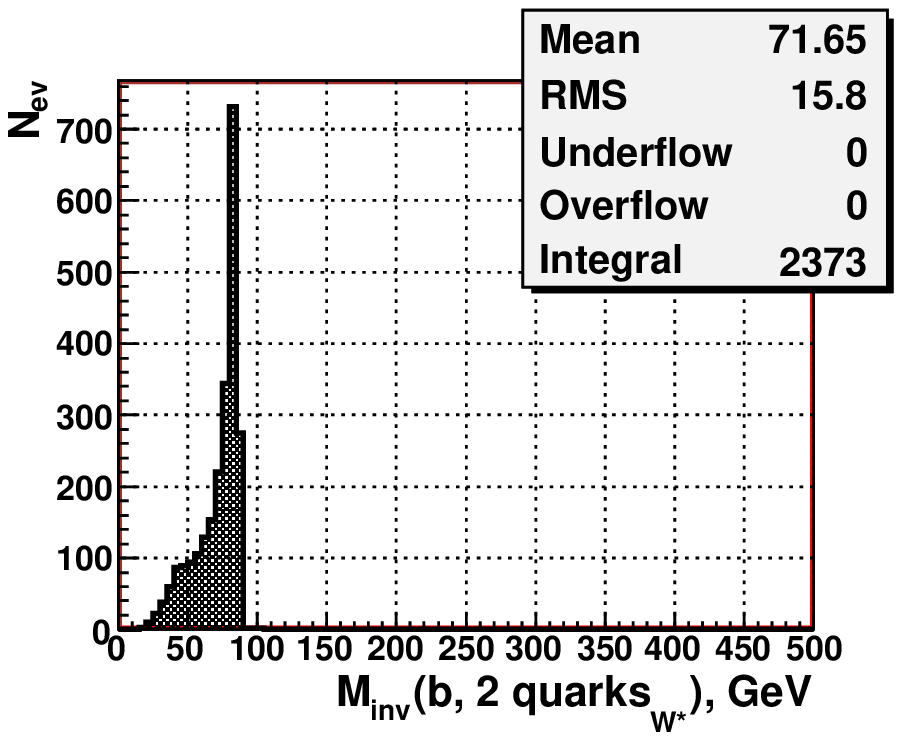}}     
      \mbox{b) \includegraphics[width=7.2cm, height=5.2cm]{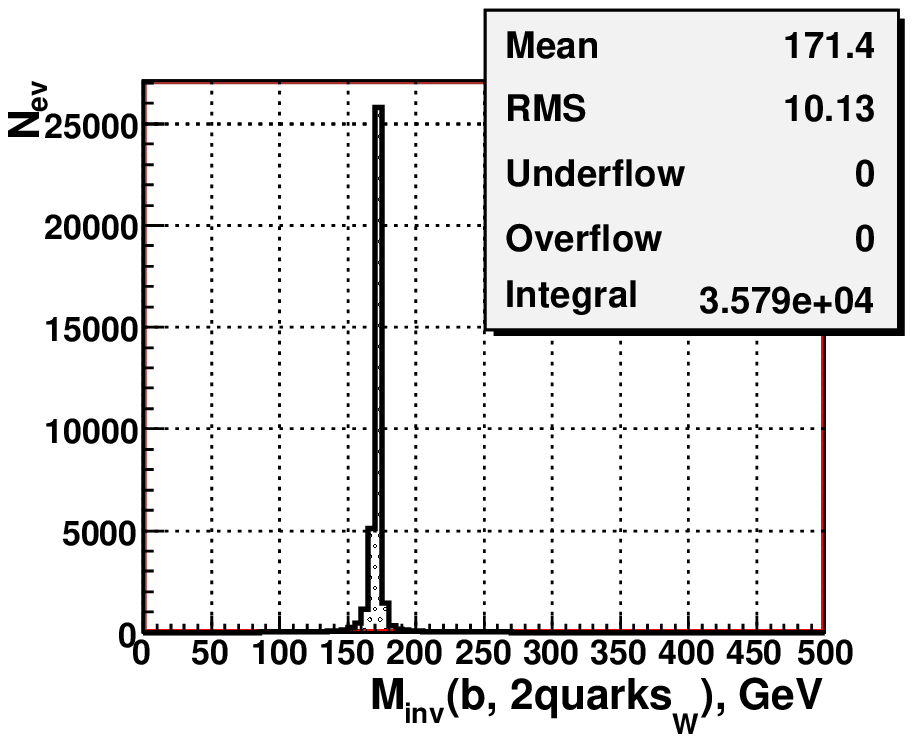}}      
   \end{tabular}
     \caption{\small \it 
                     The spectra of the invariant masses
                     $M_{inv}$($b$, $2~quarks_{W}$)
                     before the cuts (14) - (16).
	             {\bf a)}  stop pair  production;
	             {\bf b)} top  pair production.}     
     \end{center} 
\vskip -0.5cm             
    \end{figure} 
          
   In the top case the invariant mass 
   $M_{inv}(b, 2~quarks_W)$  of the system 
   composed of a $b$-quark and two quarks from
   W decay should  reproduce the mass of
   their parent top quark (see Fig.1).
   The distributions of 
   events    $dN^{event}/dM_{inv}/5$ GeV  
   expected  in each bin of 5 GeV  
   versus the  invariant mass  
   $M_{inv}(b, 2~quarks_{W})$ of 
   the parent three quarks as well as the 
   invariant mass of jets produced by these
   quarks, i.e.  $M_{inv}(b-jet, JETS_{W})$, 
   are shown  for jet  and quark  levels in plots
   {\bf b)} of Fig.23 and 24,  respectively,
   for an integrated luminosity of 1000 $fb^{-1}$.
   These distributions   have an important
   common feature. Namely, they show 
   that the peak positions at quark level 
   and at jet level, practically coincide 
   to a good accuracy    with each other
   as  well as  with the  input value of 
   the top quark mass 
   $M_{top}=170.9 (\pm 1.8)$ GeV.  
   It is also seen from  plot {\bf b)}
   of Fig.23 that the quark hadronization
   into jets leads to a broadening   of
   very small tails which are seen in  
   the invariant   mass distribution at
   quark level (plot {\bf b)} of Fig.24).
    The right tail, which appeared at jet
   level (see plot {\bf b)}) of Fig.23) 
   is a bit lower than the left tail and 
   is longer than the left one. One may
   say that the peak  shape 
   at jet level still looks more or less     symmetric.
   The main message from these plots is 
   that the appearance of  tails due to 
   quark fragmentation into jets does not
   change the position of the distribution
   peak, which allows us to reconstruct 
   the input top mass both at quark and 
   jet level{\bf s}. 

   An analogous stability
   of the peak position at the jet and quark levels
  for the stop case can be seen in the plots {\bf a)} of
   Figs.23  and 24. Note that, according to 
   the  stop decay chain (2), the 
   right edge of the peak of the invariant
   mass distribution  of the 
   "$b + 2 quarks_{W}$" system  (see 
   plot {\bf a)} of Fig.24)
   corresponds to the mass difference 
   $M_{\widetilde{t}_1}-
      M_{\tilde \chi^{0}_{1}}$.

   Now let us recall that according to
  subsection 6.1 the application of the
  cuts (14)--(16) leaves only 12 background
  top events and saves about 76$\%$ of signal
  stop events. It means that the picture
  shown in plot {\bf b)} of Fig.23 would change
  drastically and resemble a random 
  distribution of the twelve top events in a
  rather wide interval 
   \footnote{ for more details see 
              Section 8 }.
  In the case of stop pair production the distributions of the invariant mass of  the "$b + 2 quarks_{W}$"  at quark level
   and  of the system "$b$-jet+$JETS_W$" at jet
  level  system  are shown in Fig.25
 taking only those stop  events 
  that have passed the cuts (14)--(16).  

     \begin{figure}[!ht]
     \begin{center}
    \begin{tabular}{cc}
     \mbox{a) \includegraphics[  width=7.2cm, height=5.2cm]{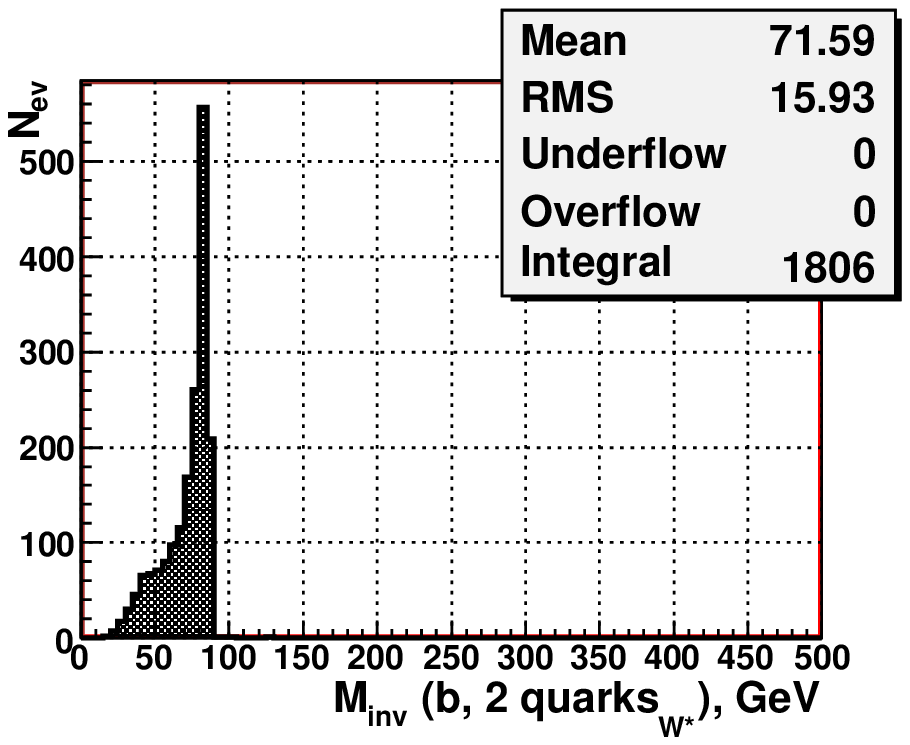}}  
     \mbox{b) \includegraphics[width=7.2cm,
     height=5.2cm]{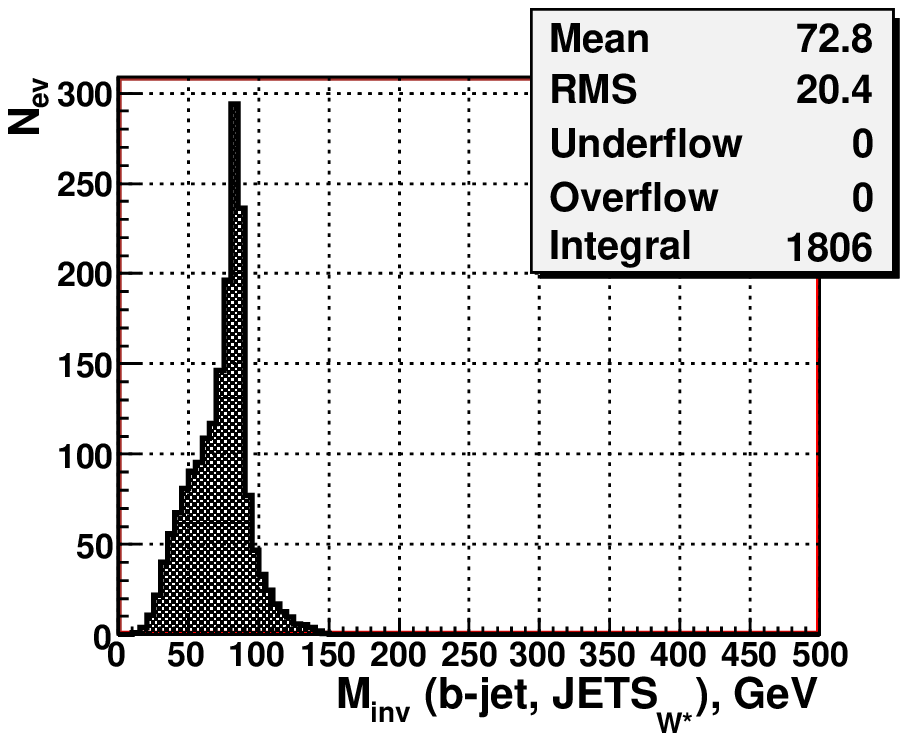}}  \\
            \end{tabular}
     \caption{\small \it The  spectra of the  stop
            signal events ($(L_{int}=1000 fb^{-1})$)
       after the cuts (14) - (16), versus 
       the invariant mass
       $M_{inv}$($b$-jet, $ JETS_{W^*}$).
	               {\bf a)} at quark level, 
	               {\bf b)} at jet level.}
  \end{center} 
\vskip -0.5cm             
    \end{figure}

  It is seen that the peak position of the
   stop distribution at jet level
   $M_{inv}(b$-jet, $ JETS_{W^{*}})$,
   obtained after application of the cuts (14)--(16) 
   (plot {\bf b)} of Fig.25), coincides with the peak
   position  at quark level (plot {\bf a)} of Fig.25) 
   as well as with the peak
   positions  in plots {\bf a)} of 
   Figs.23 and 24 obtained without any cuts.
   Let us note  that the observed  stability of the 
   peak position in both plots of Fig.25  is  due to
   the rather moderate  loss of the number
   of events in the peak region (they change 
   from $\approx 550$ to 
   $\approx 290$) while passing from quark level
   to  jet level.   The interval 150--350 GeV in the 
   plot {\bf b)} of Fig.25  can be used to calculate
   the width between the grid dots in this plot. It
   is found to be about 7.4 GeV. This number allows to
   estimate the position of the right edge of the
   peak  of $M_{inv}(b$-jet, $ JETS_{W^{*}})$
   distribution, which seems to be shifted to the
   left side from 100 GeV point by the distance
   which is a bit less than
   two dot intervals, i.e., by a bit less than 
   14.8 GeV. Thus, we can estimate that the right
   edge of the $M_{inv}(b$-jet, $ JETS_{W^{*}})$
   distribution peak lies a bit higher than 85.2 GeV.

    Some remarks about the tails in the stop distributions are 
   in order now.
   The origin  of the right and left tails
   of the  distribution shown 
   in the   plot  {\bf a)} of Fig.25 can
   be clarified  by the
   results  of the stop mass reconstruction by 
   calculating its invariant mass at quark level
   $M_{inv}$($b$, $2~quarks_{W^*}, \chi^{0}_{1})$
   as the modulus of  the sum of the  4-momenta of
   all three quarks and the neutralino (see Fig.1)
    in stop decay. These results are given in
   plot  {\bf a)} of Fig.26 which shows a very
   precise reconstruction of the input stop mass
   at quark level withing the 5 GeV width of the 
    bin containing  the peak.
   Comparing plot {\bf a)} of Fig.25 with 
   plot  {\bf a)} of Fig.26    one 
    can conclude that  at quark level 
    the long left tail as well 
    as the very small right  tail in the 
    distribution of
    $M_{inv}$($b$, $2~quarks_{W^*})$ 
   disappear when     neutralino  4-momentum is added 
to the 4-momentum of the "$b + 2 quarks_{W}$" system.

\begin{figure}[!ht]
     \begin{center}
    \begin{tabular}{cc}
     \mbox{a) \includegraphics[   width=7.2cm, height=5.2cm]{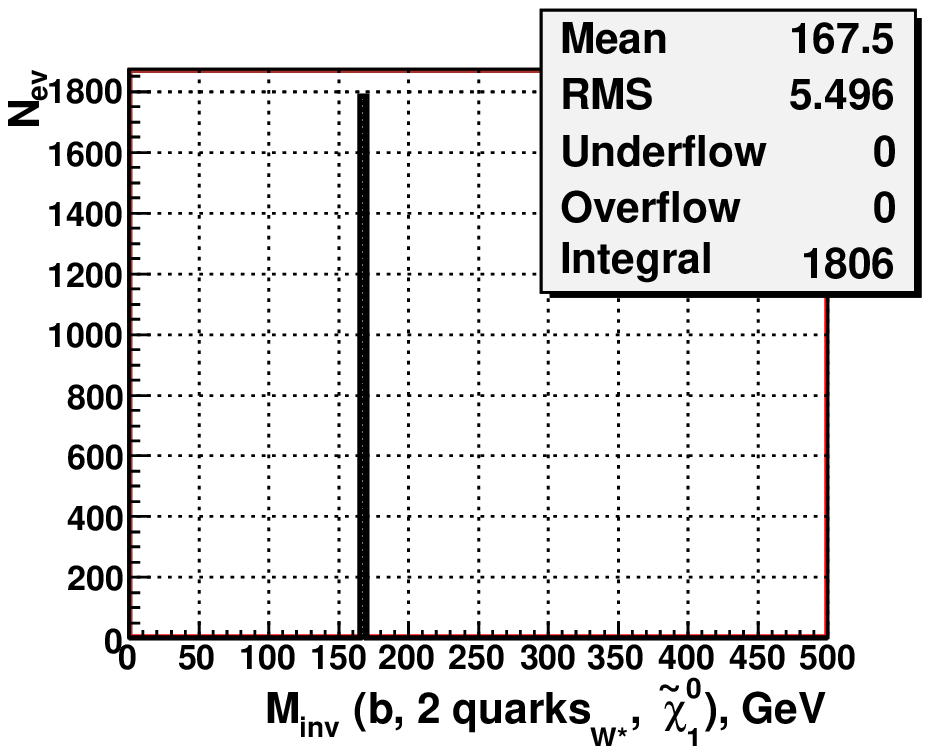}}     
    \mbox{b) \includegraphics[ width=7.2cm, height=5.2cm]{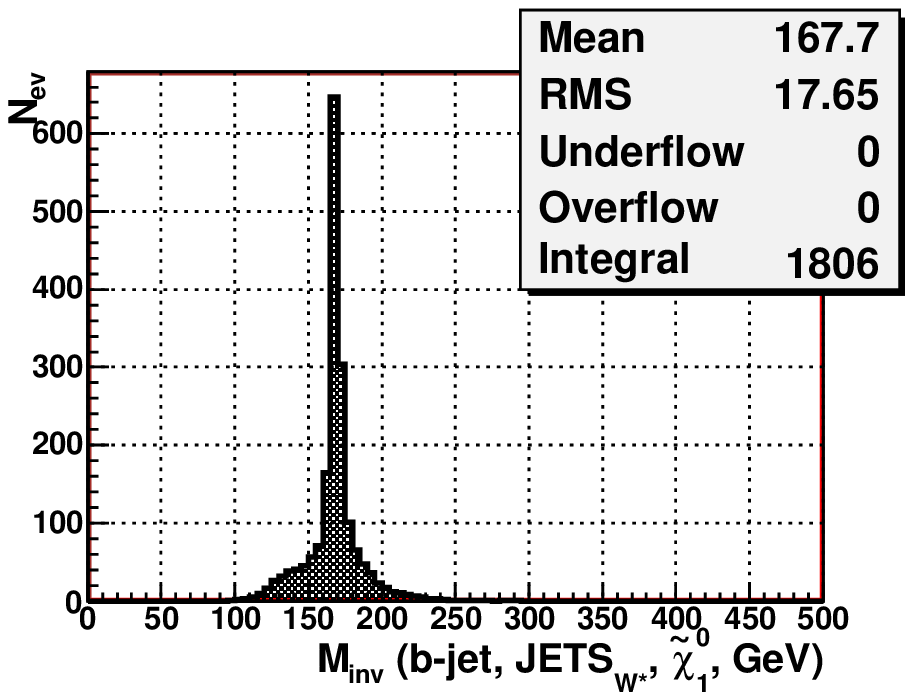}}      
    \end{tabular}
     \caption{\small \it The  spectra of the 
               stop signal events after cuts
        ($(L_{int}=1000 fb^{-1})$) versus the 
        invariant mass $M_{inv}$($b$-jet, $
        JETS_{W^*}$, $\widetilde\chi^0_1$).
	               {\bf a)} at quark level
	               {\bf b)} at jet level.}
     \end{center} 
\vskip -0.5cm             
    \end{figure}

   The  influence of the effect of  the
   hadronization of the $b$-quarks and  
   of the quarks from W decay  into jets 
   is shown in  plot  {\bf b)} of 
   Fig.26.  This plot demonstrates  
   that the hadronization of quarks into jets 
   does not  change the position of the 
   stop mass peak, which practically
   coincides with the input value  
   $M_{\widetilde{t}_1}=167.9$ GeV.
  It is also seen that the hadronization 
   results in the appearance of  more 
   or less symmetrical
   and rather  suppressed short tails 
   around the peak position.  The mean
   value of the distribution in   plot
   {\bf b)} of Fig.26 is also very
   slightly different from the 
   mean value of the quark level
   distribution shown in plot {\bf a)} 
   of Fig.26. Both of these mean values
   are in a good agreement with the 
   input value of the stop mass. 
   It is worth mentioning that  the
   shape of the peak  in  the stop
   plot {\bf b)} of Fig.26  looks 
   similar to the shape of the peak 
   in the top plot {\bf b)} of Fig.23
   which also demonstrates the  stability
    (compaired to the quark level
   top plot {\bf b)} of Fig.24)  of the 
   reconstructed top mass peak position after 
   taking into account    the effect of quark
    fragmentation to hadrons. 

  It is seen from  plot {\bf b)} of Fig.25
  that by adding the  mass of the neutralino
  $M_{\tilde \chi^{0}_{1}} = 80.9$ GeV to
  the value of the right edge point of the peak
  $M_{inv}$($b$-jet, $JETS_{W}) \approx$ 85.2 GeV
  one can get the  lower limit for the 
  reconstructed stop 
  mass $M^{reco-low}_{\widetilde{t}_1} 
                        \approx 166.2$~GeV 
  which reproduces well  the input value 
  $M_{\widetilde{t}_1}=167.9$ GeV.

  The simulation has shown 
 (see plot {\bf b)} of Fig.27) 
 that the  12 background top events, which have
  passed our cuts (14)-(16), as discussed above, are mostly
  distributed  in the region 
 $30 \leq M_{inv}$($b$-jet, $ JETS_{W}$)$ \leq 140$  GeV.
  This region is by  more than twenty 
  times  wider than 
  the  5 GeV width  of the  peak  interval in the
  $M_{inv}$($b$-jet, $ JETS_{W}$) distribution
  which is shown in  plot {\bf b)} of Fig.25 for the stop
  and  which contains  about  290 signal stop events 
  left  after the cuts. Therefore, we expect
  that  in future measurements
  the contribution of  these twelve  remaining top
  background events  will not influence 
  the  position of the peak of the
  $M_{inv}$($b$-jet, $ JETS_{W}$) distribution
  (shown in plot {\bf b)} of Fig.25) which 
  allows one to reconstruct the input  value of the
  stop  mass by adding the mass of the neutralino.
 
%
    \section{~ Results for top squark mass 
              $M_{\widetilde t_1}$ = 200 GeV.}
%

 ~~~~ In this section we want to discuss  what will 
  change if the mass of the top squark
  is different from the one we have chosen.
  In the present paper we have chosen a 
  rather low scalar top quark mass
  (one of the lowest stop quark's masses 
  that is  allowed for the case of
  $\tilde t_{1} \to b \tilde \chi_{1}^{\pm}$
  decay channel). With increase of the 
  stop mass the cross section
  for its production is
 decreasing. So, for example, for the
 case of $M_{\widetilde{t}_1}=200$ GeV 
  and the integrated luminosity
 $L_{int}=1000 fb^{-1}$  the number
 of events  per year  at $\sqrt{s}=$500 GeV is
 decreasing to 509 (after the cuts (14)--(16)). 
 The mass   $M_{\widetilde{t}_1}=200$ GeV
 is still below the  highest allowed 
  stop mass for the 
 $\tilde t_{1} \to b \tilde \chi_{1}^{\pm}$
 decay channel (which is about 255 GeV)  corresponding to 
 $M_{\chi^{+}_{1}}=159.2$ GeV 
 and  $M_{\chi^{0}_{1}}=80.9$ GeV. For
 stop masses below and above the described 
 region, the stop will decay to other channels
 which we do not consider in this paper.
   The distribution of the  invariant 
   mass $M_{inv}$($b-jet$, $JETS_{W}$)
   of the  "$b$-jet+$JETS_{W^*}$"  system 
   for events which have passed the same cuts
   (14)--(16) is shown in Fig.27. Plot  
   {\bf a)} is for stop production with
   $M_{\widetilde{t}_1}=200$ GeV, 
   plot  {\bf b)} is for top production.
    The top background also remains the 
   same  as it was given in Table 2 for 
   $\sqrt{s}=$500 GeV, i.e., about 12 events.

   \begin{figure}[!ht]
     \begin{center}
    \begin{tabular}{cc}
     \mbox{a) \includegraphics[  width=7.2cm, height=5.2cm]{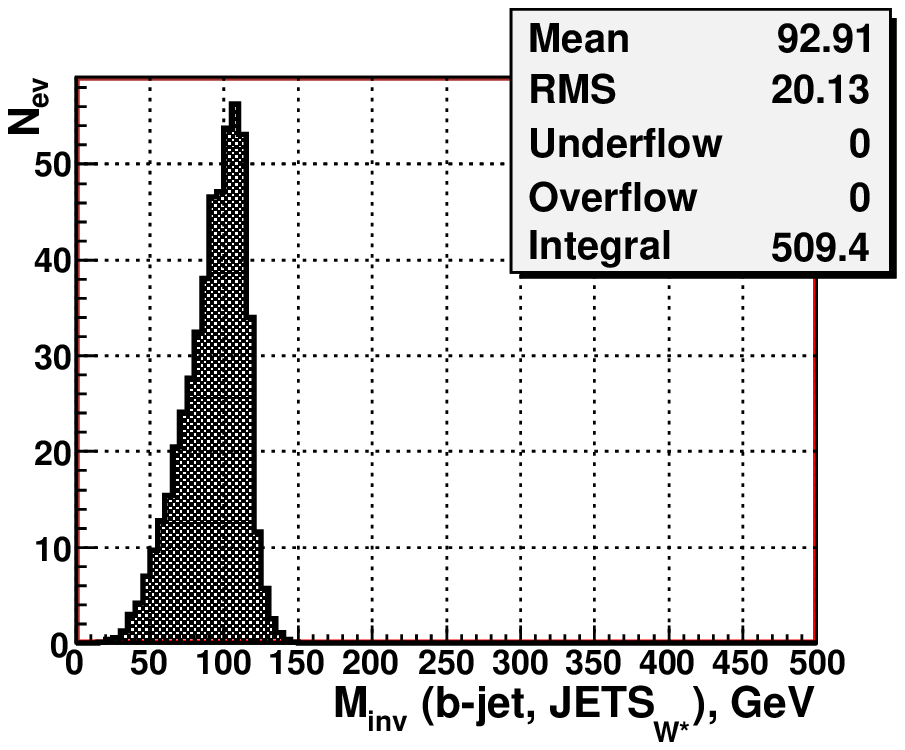}}      
     \mbox{b) \includegraphics[ width=7.2cm, 
height=5.2cm]{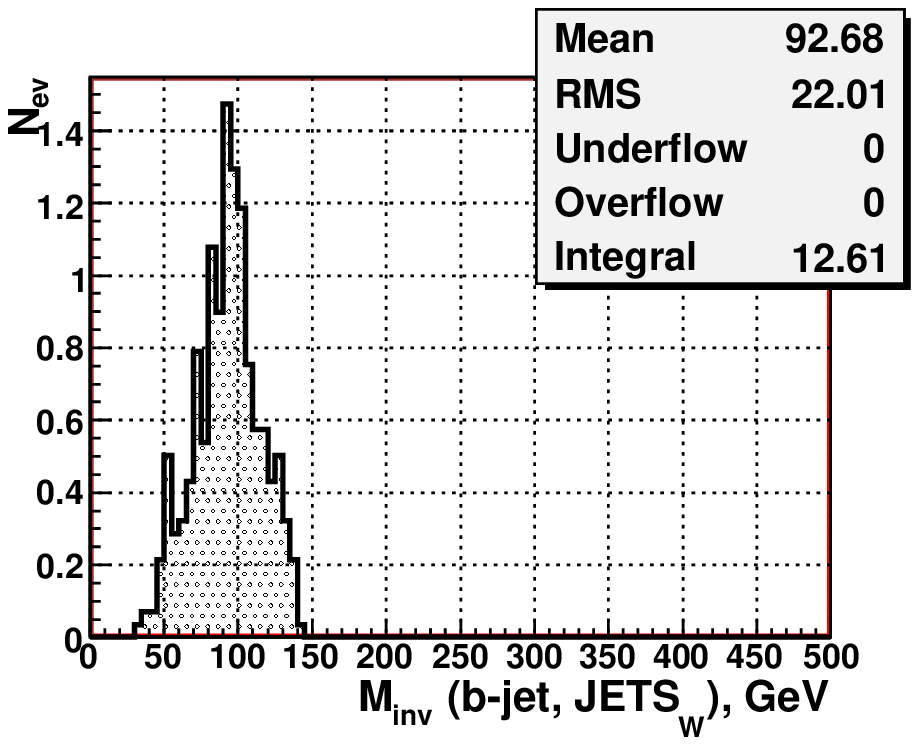}}      
    \end{tabular}
     \caption{\small \it The spectra of the   invariant masses
	  $M_{inv}$($b_{jet}$,$JETS_{W}$) of the
	 "$b$-jet+(all- non-$b$-jets)" system after cuts
	   ($L_{int}=1000 fb^{-1}$).
	               {\bf a)}  stop pair  production;
	               {\bf b)} top  pair production.}
     \end{center} 
\vskip -0.5cm             
    \end{figure}            
  The distribution  in plot {\bf a)} has a peak 
   at $M_{inv}$($b-{jet}$,$JETS_{W^{*}})\approx$
   110 GeV.  One can also determine the mass
   of the stop quark following the procedure
   described in Section  7, but with  less accuracy 
   than in the case of the lower stop mass used in 
   previous Sections.

\section{Conclusion.}

~~~~   We have studied  stop pair production 
   in electron-positron collisions within the
   framework of the  MSSM for the total energies
   $\sqrt {s}$ = 350, 400, 500, 800, 1000 GeV. 
   We assume that  the stop  quark   decays 
   dominantly into a chargino and a  $b$-quark,
   $\tilde t_{1} \to b \tilde \chi_{1}^{\pm}$, 
   and the  chargino decays  into a neutralino and 
   a  W boson,
   $\tilde \chi_{1}^{\pm} \to  \tilde \chi_{1}^{0} W^{\pm}$,
   where the W boson is virtual. One of the two
   W's decays hadronically,
   $W^{+} \to q \bar q $, the other one 
   decays leptonically, $W^{-} \to \mu^{-} \nu$.
     
     We have performed a detailed study 
   based  on a Monte Carlo simulation
   with the program  PYTHIA6.4 for 
    $E_{e+e-}^{tot}=\sqrt {s}=500$ GeV     
    and an integrated luminosity  1000 fb$^{-1}$.
At this energy we expect the highest 
    number of the signal events for 
    the chosen stop mass 
    $M_{\tilde t_{1}}$ = 167.9 GeV.
   The program CIRCE1 is used to get the
   spectra of electron  (positron)  beams 
   taking into account the effects of
   beamstrahlung.
   PYTHIA6.4  is used  to simulate 
   stop pair production and decay as
   well as top pair
   production being the main background.

    Three cuts  (14)-(16) 
    have been proposed 
    to separate  signal stop events and  top 
    background events.  For  $\sqrt {s}=500$ GeV
    and  the   luminosity  1000 fb$^{-1}$ they
     give 1806 signal stop events   with
    12  background top events.
   This is different from the more complicated 
   situation  in stop pair production at LHC 
  (see, for instance, \cite{U.Dydak}).

   We have shown that the determination  
   of the right edge of the peak position
   of the invariant  mass  
   $M_{inv}$($b$-jet,$JETS_{W^{*}}$)
   distribution of the 
   "$b$-jet+(all-non-$b$-jets)" system 
   allows us to  measure  the mass of the
   stop quark with a good  accuracy based 
   on the statistics corresponding to
  an integrated  luminosity of 1000 fb$^{-1}$. 
   For this the mass of $\chi_{1}^{0}$ has 
   to be known.

   As seen from the Table 2
   the measurements at other energies in the 
   regions  $400 \leq \sqrt {s} \leq 500$ GeV
   and  $500 \leq \sqrt {s} \leq 800$ GeV
   may allow to enlarge
   substantially the  number of selected 
   signal stop events and to perform a
   precise measurement  of the
   mass of the scalar top quark. 
  
   In the last Section we discussed the
   difference in  the main invariant distribution 
  for a  $M_{\tilde t_{1}}$ = 200 GeV.
  
    In conclusion  we can say that the $e^{+}e^{-}$
  channel together with the   $\gamma \gamma$ channel
  considered in our  previous Note \cite{BMMSS-STOP} 
  is  well suited for the study of stop pair 
  production at  ILC. 

\section{Acknowledgements.}

~~~ This work is supported by the JINR-BMBF 
 project and 
 by the "Fonds zur F$\ddot o$rderung der
 wissenschaftlichen Forschung" (FWF) of
 Austria, project No.P18959-N16.
 The authors acknowledge support from 
 EU under the MRTN-CT-2006-035505 
 and MRTN-CT-2004-503369 network programmes.
 A.B. was supported by the Spanish 
 grants SAB 2006-0072, FPA 2005-01269
 and FPA 2005-25348-E of the
 Ministero de Educacion y Ciencia.


\begin{thebibliography}{99}

\bibitem{SUSY}{Y.Gol'fand and E.Likhtman, JETP Lett. 13(1971)~323;\\
            D.Volkov and V.Akulov, Phys.Lett.
            B46(1973)~109;\\
	    J.Wess and B.Zumino, Nucl.Phys.
            B70(1974)~39.} 

\bibitem{JEllis}{J.Ellis and S.Rudaz, Phys.Lett. B128(1983)~248.}
 
\bibitem{STOP_SUSY}{G.Altarelli and R.R$\ddot u$ckl, Phys.Lett. B144(1984)~126;\\
	      S.Dawson, E.Eichten and C.Quigg, Phys.Rev.
	       D31(1985)~1581;\\
	      K.Hikasa and M.Kobayashi, Phys.Rev.
              D36(1987)~742;\\
	      M.Drees and K.Hikasa, Phys.Lett.
              B252(1990)~127;\\
	      J.Ellis, G.L.Fogli and E.Lisi, 
	      Nucl.Phys. B393(1993)~3.}

		
\bibitem{ILCRDR1}{ILC Reference Design Report, v.1 
                      "Executive Summary",\\
                      Editors: J.Brau, Y.Okada,
                      N.Walker, 2007;\\
 http://www.linearcollider.org/cms/?pid=1000025.}
		      
\bibitem{ILCRDR2}{ILC Reference Design Report, 
                      v.2 "Physics at the ILC",\\
                      Editors: A.Djouadi, J.Lykken, 
		      K.M$\ddot o$nig, Y.Okada,
                      M.Oreglia, S.Yamashita, 2007; 
       http://www.linearcollider.org/cms/?pid=1000025.}


\bibitem{BMMSS-STOP}{A.Bartl, K.M$\ddot o$nig, W.Majerotto, A.Skachkova, N.Skachkov,\\
             "Pair Production of Scalar Top Quarks
             in Polarized
	      Photon-Photon Collisions at ILC",
              ILC-NOTE-2007-036,
	      arXiv:0804.1700[hep-ph].}
 
\bibitem{A. Bartl}{A.Bartl, H.Eberl, S.Kraml,
                W.Majerotto and  W.Porod,
                Eur.Phys.J.C2(2000)6; \\ hep-ph/0002115.} 

\bibitem{Paris2004}{A.Bartl, K.Moenig, 
            W.Majerotto, A.Skachkova, N.Skachkov,\\
            ''Stop pair production in polarized
            photon-photon collisions'',\\
	    Proc. of the
            Intern. Conf. on  Linear Colliders
            (LCWS 2004), vol.II, p.919,
            April 19-23, 2004, 
            Le Carre des Sciences, Paris, France.}
                   
\bibitem{H.Nowak1}{ A.Finch, H.Nowak and
                  A.Sopczak, hep-ph/0211140.} 

\bibitem{H.Nowak2}{ M.Carena et.al, Phys.Rev.
                   D72:115008, 2005; hep-ph/0508152.} 

\bibitem{H.Nowak3}{ A. Sopczak  et.al,  hep-ph/0605225.}

\bibitem{T. Sjostrand94}{  T. Sj$\ddot o$strand, S. Mrenna and  P.Skands
         JHEP 0605:026, 2006; hep-ph/0603175v2

\bibitem{T.Ohl}{ T. Ohl, 
   "$\it {\kappa \iota \rho \kappa \eta}$ Version 1.0:
    Beam Spectra for Simulating Linear Collider
    Physics", hep-ph/9607454. }

\bibitem{Gunion}{J.F.Gunion, H.E.Haber,
        Nucl.Phys. B272(1986)1;
        B278(1986)449; Erratum B402(1993)567.}

\bibitem{Schieferdecker}{E. Brubaker et al.   "Combination of CDF and D0 results          on the mass of the top quark";
        By Tevatron Electroweak Working Group,
        Fermilab-TM-2380-E,
        19 Mar2007; arXiv:hep-ex/0703034.}
		     

\bibitem{Durham}{S.Catani, Yu.L. Dokshitzer, M.Olson, G.Turnock
	 and B.Webber, Phys.Lett. B269 (1991) 432.}
 
\bibitem{JADE}{JADE Collab., W.Bartel et al. Z.Phys. C33 (1986) 23;
               S.Bethke, Habilitation thesis, 
                         LBL, 50-208, 1987.}

\bibitem{MoenigKlamke}{G.Klamke and K.Moenig, 
Eur.Phys.J.C42(2005)261, DESY-05-049, Mar 2005, hep-ph/0503191.}  

\bibitem{Haw}{R.Hawkings, "Vertex detector and flavour tagging studies for
TESLA liear collider", LC-PHSM-2000-021, 2000}

\bibitem{U.Dydak}{ U.Dydak, "Search for the stop quark with CMS at
              the LHC";  CMS TN/96-022, CERN, 1996; \\ 
              U.Dydak, H.Rohringer
              and J.Tuominiemi, "Study of the channel 
              gluino $\to$ stop + top", 
              CMS TN/96--103, CERN, 1996.}


\bibitem{Damerell}{C.J.S.Damerell, D.J.Jackson, eConf960625 (1996) DET078;\\
R.Hawkings, LC-PHSM-2000-021; \\
S.M.Xella Hansen, D.J.Jackson, R.Hawkings, C.J.S.Damerell,
LC-PHSM-2001-024; \\
S.M.Xella Hansen, M.Wing, D.J.Jackson, N. De Groot, C.J.S.Damerell,
LC-PHSM-2003-061; \\
S.M.Xella Hansen et al. [Linear Collider Flavour Identification Collaboration], Nucl.Instrum.Meth.A501,106(2003);
S.Hillert, C.J.S.Damerell, eConf0508141 (2005) ALCPG 1403.}

\bibitem{Gudi}{G.Moortgat-Pick et.al., "The role of polarized positrons 
         and  electrons in revealing fundamental
	 interactions at the Linear Collider",
	 Phys.Rept.460 (2008) 131;
         hep-ph/0507011.}
 
\bibitem{DShulte}{D. Schulte, "Study of electromagnetic and hadronic 
	              background in the interaction region 
		      of the TESLA Collider, 
		      Ph.D. Thesis, Univ. Hamburg, 1997,
		      DESY-TESLA-97-08;
		      	 
	 K.Yokoya 
         and P.Chen, KEK Preprint 91-2, April 1991;\\
         http://www-sldnt.slac.stanford.edu/
	 nlc/programs/guinea$\_$pig/gp$\_$index.html.}
	 
 
} 

\end{thebibliography}
\end{document}